\documentclass{ifacconf}

\usepackage[english]{babel}
\usepackage{amssymb}
\usepackage{stmaryrd}
\usepackage{tikz}
\usepackage{mathtools}
\usepackage{etoolbox}
\usepackage[ruled,vlined,algo2e,linesnumbered]{algorithm2e}
\usepackage{nicematrix}

\DeclareMathAlphabet{\pazocal}{OMS}{zplm}{m}{n}
\usepackage [autostyle, english = american]{csquotes} 
\MakeOuterQuote{"}

\usepackage{xargs}
\usepackage{soul, color, xcolor, graphicx}
\usepackage{natbib}

\usetikzlibrary{positioning,arrows,petri,calc,decorations.markings,arrows.meta}
\tikzset{
place/.style={circle,thick,minimum size=4mm,draw},
transitionV/.style={rectangle,thick,fill=black,minimum height=6mm,inner xsep=1pt}
}

\definecolor{myblue}{RGB}{0, 101, 202}
\definecolor{mygreen}{RGB}{130, 180, 0}
\definecolor{myred}{RGB}{197, 14, 31}
\definecolor{mypurple}{RGB}{128, 0, 128}
\definecolor{myyellow}{RGB}{204, 204, 0}
\definecolor{mygrey}{RGB}{105, 105, 105}

\newcommand{\dint}[1]{\left\llbracket#1\right\rrbracket} % discrete interval

\newcommand{\D}{\pazocal{D}}
\renewcommand{\S}{\pazocal{S}}

\newcommand{\N}{\mathbb{N}}
\newcommand{\No}{\mathbb{N}_0}
\newcommand{\Z}{\mathbb{Z}}

\newcommand{\R}{\mathbb{R}}
\newcommand{\Q}{\mathbb{Q}}

\newcommand{\Qmax}{{\Q}_{\normalfont\fontsize{7pt}{11pt}\selectfont\mbox{max}}}
\newcommand{\Qmin}{{\Q}_{\normalfont\fontsize{7pt}{11pt}\selectfont\mbox{min}}}
\newcommand{\Qbar}{\overline{\Q}}

\newcommand{\zor}[1]{{\color{red}#1}}
\newcommand{\rewrite}[2]{#1} % toggle this from #1 to #2 to display only the second version
\newcommand{\rerewrite}[2]{#1} % toggle this from #1 to #2 to display only the second version

\newcommand{\graph}{G}
\newcommand{\places}{\pazocal{P}}
\newcommand{\transitions}{\pazocal{T}}
\newcommand{\arcs}{E}
\newcommand{\nodes}{N}
\newcommand{\shift}{\mbox{shift}}
\newcommand{\weight}{\mbox{weight}}
\renewcommand{\lshift}{\mbox{Lshift}}

\newcommand{\up}{\mbox{source}}
\newcommand{\down}{\mbox{target}}
\newcommand{\height}{\mbox{base}}
\newcommand{\len}{\mbox{len}}

\makeatletter
\newcommand{\splus}{%
  \DOTSB\mathop{\mathpalette\mattos@splus\relax}\slimits@
}
\newcommand\mattos@splus[2]{%
  \vcenter{\hbox{%
    \sbox\z@{$#1\oplus$}%
    \resizebox{!}{0.9\dimexpr\ht\z@+\dp\z@}{\raisebox{\depth}{$\m@th#1\boxplus$}}%
  }}%
  \vphantom{\oplus}%
}
\makeatother

\makeatletter
\newcommand{\stimes}{%
  \DOTSB\mathop{\mathpalette\mattos@stimes\relax}\slimits@
}
\newcommand\mattos@stimes[2]{%
  \vcenter{\hbox{%
    \sbox\z@{$#1\otimes$}%
    \resizebox{!}{0.9\dimexpr\ht\z@+\dp\z@}{\raisebox{\depth}{$\m@th#1\boxtimes$}}%
  }}%
  \vphantom{\otimes}%
}
\makeatother

\usepackage{pict2e}

\makeatletter
\newcommand*{\bigsplus}{\DOTSB\mathop{\mathpalette\big@boxplus\relax}\slimits@}

\newcommand{\big@boxplus}[2]{%
  \vcenter{%
    \m@th\bigbox@thickness{#1}%
    \sbox\z@{$#1\bigoplus$}%
    \dimen@=\ht\z@ \advance\dimen@\dp\z@
    \hbox{%
      \setlength{\unitlength}{\dimen@}%
      \begin{picture}(1,1)
      \polyline(0.1,0.1)(0.9,0.1)(0.9,0.9)(0.1,0.9)(0.1,0.1)(0.5,0.1)
      \polyline(0.5,0.1)(0.5,0.9)
      \polyline(0.1,0.5)(0.9,0.5)
      \end{picture}%
    }%
  }%
}

\newcommand{\bigbox@thickness}[1]{%
  \ifx#1\displaystyle
    \linethickness{0.2ex}%
  \else
    \ifx#1\textstyle
      \linethickness{0.16ex}%
    \else
      \ifx#1\scriptstyle
        \linethickness{0.12ex}%
      \else
        \linethickness{0.1ex}%
      \fi
    \fi
  \fi
}
\makeatother
\newcommand{\svdots}{\raisebox{3pt}{$\scalebox{.75}{\vdots}$}} 
\newcommand{\sddots}{\raisebox{3pt}{$\scalebox{.75}{$\ddots$}$}} 

\newcommand*{\myendproofsymbol}[1][$\blacksquare$]{%
\leavevmode\unskip\penalty9999 \hbox{}\nobreak\hfill
    \quad\hbox{#1}%
}
\newcommand*{\myendtheoremsymbol}[1][$\square$]{%
\leavevmode\unskip\penalty9999 \hbox{}\nobreak\hfill
    \quad\hbox{#1}%
}
\newcommand*{\myenddefinitionsymbol}[1][$\lozenge$]{%
\leavevmode\unskip\penalty9999 \hbox{}\nobreak\hfill
    \quad\hbox{#1}%
}

\begin{document}

% double spacing for Joerg:
%\linespread{2}
%\openup 1em

\begin{frontmatter}

\title{Consistency of P-time event graphs is decidable in polynomial time\rewrite{\\(extended version)}{}{}} 

\author[First]{Davide Zorzenon} 
\author[First,Second]{J\"{o}rg Raisch\thanksref{footnoteinfo}} 

\thanks[footnoteinfo]{This work was funded by the Deutsche Forschungsgemeinschaft (DFG, German Research Foundation), Projektnummer RA 516/14-1.
Partially supported by the Deutsche Forschungsgemeinschaft (DFG, German Research Foundation), under Germany's Excellence Strategy - EXC 2002/1 "Science of Intelligence" - project number 390523135.}

\address[First]{Control Systems Group, Technische Universit\"{a}t Berlin, Germany 
  (email: [zorzenon,raisch]@control.tu-berlin.de).}
\address[Second]{Science of Intelligence, Research Cluster of Excellence, Berlin, Germany.}

\begin{abstract}
P-time event graphs are discrete event systems able to model cyclic production systems where tasks need to be performed within given time windows.
Consistency is the property of admitting an infinite execution of such tasks that does not violate any temporal constraints.
In this paper, we solve the long-standing problem of characterizing the decidability of consistency by showing that, assuming unary encoding of the initial marking, this property can be verified in strongly polynomial time.
The proof is based on a reduction to the problem of detecting paths with infinite weight in infinite weighted digraphs called $\N$-periodic graphs.
\end{abstract}


\end{frontmatter}
%===============================================================================

\section{Introduction}

In many production systems, ranging from the food industry to PCB manufacturing, the violation of temporal specifications can result in irreparable damage of the final product. 
When the sequence of logical operations to be performed is infinite and cyclically repeating, such systems can be modeled by P-time event graphs (P-TEGs).
P-TEGs are ordinary Petri nets where time intervals are associated to places, and each place has exactly one upstream and one downstream transition.
Failure to meet a temporal specification in the real system corresponds to a token remaining in a place of the P-TEG for longer than prescribed by the associated time-window constraint.

Despite their simple -- albeit nondeterministic -- dynamics characterized by linear inequalities, a fundamental question has remained open for decades (\cite{khansa1996p}): is it possible to verify in finite time whether a P-TEG is consistent, that is, admits an infinite sequence of firings of transitions that does not violate any constraint?
%The consistency property plays a central role in understanding production systems with time-window specifications.
This question is motivated by the fact that several problems of practical importance concerning performance evaluation and control of production systems modeled by P-TEGs cannot be solved without a thorough understanding of the consistency property.
%the open status of this question implies other interesting problems concerning the performance evaluation and control of P-TEGs are still open as well.
For instance, it is currently not known whether there exist algorithms to compute the throughput of a P-TEG\footnote{Except for the restricted, although interesting, case where the P-TEG is forced to follow periodic trajectories.}, or to find a just-in-time-optimal trajectory according to given deadlines.
%If such algorithms were available, indeed, they could be used to check the consistency property, contradicting the open status of the question.
%For instance, if an algorithm that computes the throughput of a P-TEG were available, we could use it to check consistency as consistent P-TEGs are the only ones with positive throughput.
%As a second example, having an algorithm to fire the transitions of a P-TEG according to the just-in-time policy would also contradict the status of the consistency problem.

The consistency verification problem has been considered by several authors for special cases.
Given a P-TEG with $n$ transitions and at most one initial token in each place\footnote{Any P-TEG can be transformed into another, behaviorally equivalent one where this condition is met, at the cost of increasing the number of transitions. See Remark~\ref{re:transformation} for details.}, the problem of checking the existence of acceptable trajectories of a given finite length $h$ was proven to be solvable in time $\pazocal{O}(hn^3)$ in~\cite{5628259}.
A generalization of this result is given in~\cite{ZORZENON202219}, where the concept of weak consistency was introduced.
A P-TEG is called weakly consistent if it admits trajectories of any finite length.
Interestingly, this property does not imply consistency, as some P-TEGs can admit finite trajectories of any length, but no infinite trajectory\footnote{An example of weakly consistent but not consistent P-TEG is given in Figure~\ref{fi:P-TEG_example} for parameters $\alpha=-5$, $\beta=4$.}.
%A scenario in which weak consistency is not sufficient is in continuous productions systems, 
In~\cite{ZORZENON202219}, it was shown that weak consistency can be verified in strongly polynomial time $\pazocal{O}(n^9)$.
The consistency verification problem was essentially solved for the case where upper bound constraints appear only in places with no initial tokens in~\cite[Corollary 2.3]{iteb2006control}.
Building on the same approach based on formal power series, in the PhD thesis~\cite{brunsch2014modeling}, the consistency problem was declared solved in its entirety.
However, the solution proposed is invalid due the presence of a technical error\footnote{Using the notation of~\cite{brunsch2014modeling}, the dual product $\odot$ does \textit{not} distribute over the infimum $\wedge$ in the dioid $\pazocal{M}_{in}^{ax}\dint{\gamma,\delta}$. For a counterexample, one can check that, for $a=\gamma^3\delta^1$, $b=\gamma^5\delta^3$, $c=\gamma^1\delta^2\oplus \gamma^3\delta^5$, $(a\wedge b)\odot c = \gamma^6\delta^3\oplus \gamma^8\delta^6 \neq \gamma^6\delta^5\oplus \gamma^8\delta^6 =  (a\odot c) \wedge (b \odot c)$. This, among other results, invalidates the method given in Section 5.3 of~\cite{brunsch2014modeling} to detect unfeasible constraints.}.
Other only necessary and only sufficient conditions for consistency were given, e.g., in~\cite{komenda2011application,zorzenon2020bounded,vspavcek2021analysis}.

\rerewrite{
In the present paper, we finally answer the investigated question by providing an algorithm that checks consistency in strongly polynomial time complexity $\pazocal{O}(n^6)$.
The algorithm is based on the connection between P-TEGs, infinite block-tridiagonal matrices in the max-plus algebra, and $\N$-periodic graphs, which are infinite weighted directed graphs obtained, roughly speaking, by placing a finite graph in each point of the lattice of natural numbers $\N$.
In particular, in Section~\ref{se:N_periodic_graphs}, we show that it is possible to detect the presence of paths with infinite weight in $\N$-periodic graphs in polynomial time.
We mention that some of the tools employed in the proof, such as well partial orders and Diophantine equations, were drawn from an interesting connection with vector addition systems with states (\cite{leroux2015demystifying,clementehal-01587619}).
This fact is used in Section~\ref{se:algebra} to provide an algorithm that checks if the max-plus Kleene star of an infinite block-tridiagonal matrix contains entries equal to $+\infty$.
Finally, Section~\ref{se:P_TEGs} provides the connection between P-TEGs and infinite block-tridiagonal matrices.
}{
In the present paper, we finally answer the investigated question by providing a reduction of the consistency verification problem to the problem of detecting the existence of $\infty$-weight paths in $\N$-periodic graphs, which has been shown to be polynomial-time solvable in~\zor{CITE THE ARXIV PAPER}.
$\N$-periodic graphs (defined in Section~\ref{se:N_periodic_graphs}) are infinite weighted directed graphs obtained, roughly speaking, by placing a finite graph in each point of the lattice of natural numbers $\N$.
%We mention that some of the tools employed to establish the polynomial-time detectability of $\infty$-weight paths in $\N$-periodic graphs, such as well partial orders and Diophantine equations, were drawn from an interesting connection with vector addition systems with states (\cite{leroux2015demystifying,clementehal-01587619}).
The link between P-TEGs and $\N$-periodic graphs is provided by their infinite adjacency matrices of block-tridiagonal structure, introduced in Section~\ref{se:algebra}.
%In Section~\ref{se:algebra}, we show that the max-plus Kleene star of an infinite block-tridiagonal matrix
%In Section~\ref{se:algebra}, we provide the he connection between P-TEGs and $\N$-periodic graphs is  is based on the connection between P-TEGs, infinite block-tridiagonal matrices in the max-plus algebra, and $\N$-periodic graphs, which are infinite weighted directed graphs obtained, roughly speaking, by placing a finite graph in each point of the lattice of natural numbers $\N$.
%In particular, in Section~\ref{se:N_periodic_graphs}, we show that it is possible to detect the presence of paths with infinite weight in $\N$-periodic graphs in polynomial time.
%This fact is used in Section~\ref{se:algebra} to provide an algorithm that checks if the max-plus Kleene star of an infinite block-tridiagonal matrix contains entries equal to $+\infty$.
In Section~\ref{se:P_TEGs}, it is shown that consistency of P-TEGs is equivalent to the absence of entries equal to $+\infty$ in the max-plus Kleene star of associated infinite block-tridiagonal matrices, and this establishes the reduction.
}

\rewrite{}{Due to space limitations, some of the proofs of Sections~\ref{se:N_periodic_graphs} and~\ref{se:algebra} are omitted here and presented in~\zor{CITE THE ARXIV PAPER}.}{}

\subsubsection*{Notation}
We denote sets $\Q\cup\{-\infty\}$, $\Q\cup\{\infty\}$, $\Q\cup\{\pm\infty\}$ respectively by $\Qmax$, $\Qmin$, and $\Qbar$.
The sets of nonnegative and positive integers are denoted, respectively, by $\No$ and $\N$.
Given $a,b\in\Z$, with $b\geq a$, $\dint{a,b}$ indicates the discrete interval $\{a,a+1,a+2,\ldots,b\}$.
\section{$\infty$-weight paths in $\N$-periodic graphs}\label{se:N_periodic_graphs}

In this section, we \rerewrite{study}{introduce} a decision problem related to $\N$-periodic graphs.
Let us first define static graphs. %$\S$-periodic graphs.

\begin{defn}[Static graph]
Given three $n\times n$ matrices $M_{-1}$, $M_{0}$, $M_{+1}$ with elements from $\Qmax$, the associated static graph $\graph = \graph(M_{-1},M_{0},M_{+1}) = (V,E,w)$ is the weighted multi-directed graph with set of nodes $V = \dint{1,n}$, set of arcs $E\subseteq V\times V\times \{-1,0,+1\}$, and weight function $w:E\rightarrow \Q$, defined such that there is an arc $e = (i,j,s)\in E$ from the upstream node $i \eqqcolon \up(e)$ to the downstream node $j \eqqcolon \down(e)$ with shift $s \eqqcolon \shift(e)$ and weight $(M_{s})_{ji} \eqqcolon w(e)$ if and only if $(M_{s})_{ji} \neq -\infty$.\myenddefinitionsymbol
\end{defn}

%Matrices $M^{(-1)},M^{(0)},M^{(+1)}$ will be often denoted by $L,C,R$ for left, center, and right matrices.
Every static graph induces (or generates) $\Z$-periodic, $\N$-periodic, and -- more generally -- $\S$-periodic graphs as follows.

\begin{defn}[$\S$-periodic graphs]
Let $\S$ be any subset of $\Z$ and let $G = \graph(M_{-1},M_0,M_{+1}) = (V,E,w)$ be a static graph.
The $\S$-periodic graph $\graph_\S = \graph_\S(M_{-1},M_0,M_{+1}) = (V_{\S},E_{\S},w)$ induced by $G$ is the (possibly infinite) weighted directed graph with set of nodes $V_{\S} = V \times \S$, set of arcs 
\[
    E_\S = \{((i,k),(j,k+s))\mid (i,j,s)\in E,\ k,k+s\in\S\},
\] 
and weight function\footnote{With slight abuse of notation, we denote by $w$ the weight functions of both $G$ and $G_\S$.} $w:E_\S\rightarrow \Q$ defined by $w((i,k),(j,k+s)) = w(i,j,s)$.
The base and shift of a node $v_\S = (i,k)\in V_\S$ are, respectively, $\height(v_\S) \coloneqq i$ and $\shift(v_\S) \coloneqq k$.
For every arc $e_\S = ((i,k),(j,k+s))\in E_\S$, we define $\up(e_\S) \coloneqq (i,k)$, $\down(e_\S) \coloneqq (j,k+s)$, $\shift(e_\S) \coloneqq s$.\myenddefinitionsymbol
\end{defn}

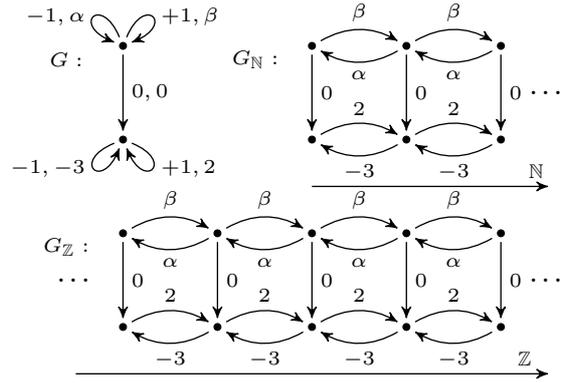
\begin{figure}
    \centering
    \resizebox{.9\linewidth}{!}{
        \begin{tikzpicture}[node distance=2cm and 2cm,>=stealth',bend angle=45,double distance=.5mm,arc/.style={->,>=stealth'},place/.style={circle,thick,minimum size=8mm,draw}]

\tiny

% G

\node (nttop) at (1,2) {};
\node (nbbot) at (1,1) {};
\filldraw (nttop) circle (1pt);
\filldraw (nbbot) circle (1pt);
\draw [arc] (nttop) to node[auto] {$0,0$} (nbbot);
\draw [arc] (nttop) to [out=45+22.5,in=45-22.5,loop] node[right] {$+1,\beta$} (nttop);
\draw [arc] (nttop) to [out=90+45-22.5,in=90+45+22.5,loop] node[left] {$-1,\alpha$} (nttop);
\draw [arc] (nbbot) to [out=-45+22.5,in=-45-22.5,loop] node[right] {$+1,2$} (nbbot);
\draw [arc] (nbbot) to [out=-90-45-22.5,in=-90-45+22.5,loop] node[left] {$-1,-3$} (nbbot);
\node at (0.4,1.85) {$\graph:$};

% G_Z

\foreach \z in {1,2,3,4,5}
{
\node (ntop\z) at (\z,0) {};
\node (nbot\z) at (\z,-1) {};
\filldraw (ntop\z) circle (1pt);
\filldraw (nbot\z) circle (1pt);
\draw [arc] (ntop\z) to node[auto] {$0$} (nbot\z);
}

\foreach[count=\i, evaluate=\i as \zz using int(\i+1)] \z in {1,2,...,4}
{
\draw [arc] (ntop\z) to [bend left=30] node[auto] {$\beta$} (ntop\zz);
\draw [arc] (nbot\z) to [bend left=30] node[auto] {$2$} (nbot\zz);
}

\foreach[count=\i, evaluate=\i as \zz using int(\i)] \z in {2,3,...,5}
{
\draw [arc] (ntop\z) to [bend left=30] node[auto] {$\alpha$} (ntop\zz);
\draw [arc] (nbot\z) to [bend left=30] node[auto] {$-3$} (nbot\zz);
}

\node (dots1) at (0.5,-.5) {\textbf{\dots}};
\node (dots2) at (5.5,-.5) {\textbf{\dots}};

\draw [arc] (.5,-1.5) to node[pos=.95,above] {$\Z$} (5.5,-1.5);
\node at (0.4,-0.15) {$\graph_\Z:$};

% G_N

\foreach \z in {3,4,5}
{
\node (ntop\z) at (\z,2) {};
\node (nbot\z) at (\z,1) {};
\filldraw (ntop\z) circle (1pt);
\filldraw (nbot\z) circle (1pt);
\draw [arc] (ntop\z) to node[auto] {$0$} (nbot\z);
}

\foreach[count=\i, evaluate=\i as \zz using int(\i+3)] \z in {3,4}
{
\draw [arc] (ntop\z) to [bend left=30] node[auto] {$\beta$} (ntop\zz);
\draw [arc] (nbot\z) to [bend left=30] node[auto] {$2$} (nbot\zz);
}

\foreach[count=\i, evaluate=\i as \zz using int(\i+2)] \z in {4,5}
{
\draw [arc] (ntop\z) to [bend left=30] node[auto] {$\alpha$} (ntop\zz);
\draw [arc] (nbot\z) to [bend left=30] node[auto] {$-3$} (nbot\zz);
}

%\node (dots1) at (0.5,-.5) {\textbf{\dots}};
\node (dots2) at (5.5,1.5) {\textbf{\dots}};

\draw [arc] (3,0.5) to node[pos=.95,above] {$\N$} (5.5,0.5);
\node at (2.4,1.85) {$\graph_\N:$};

\end{tikzpicture}
    }
    \caption{An example of static graph $\graph$ (top left), and its induced $\N$-periodic graph $\graph_\N$ (top right) and $\Z$-periodic graph $\graph_\Z$ (bottom).
    Every arc $e$ in the static graph is labeled "$\shift(e),w(e)$".}
    \label{fi:periodic_graph}
\end{figure}

\begin{exmp}\label{ex:simple_graph}
Figure~\ref{fi:periodic_graph} illustrates the static graph associated with matrices $M_{-1} = \begin{bsmallmatrix}
        \alpha & -\infty\\-\infty & -3
    \end{bsmallmatrix}$, $M_{0} = \begin{bsmallmatrix}
        -\infty & -\infty\\0 & -\infty
    \end{bsmallmatrix}$, $M_{+1} = \begin{bsmallmatrix}
        \beta & -\infty\\-\infty & 2
    \end{bsmallmatrix}$,
where $\alpha,\beta\in\Q$, together with (portions of) its induced $\N$-periodic and $\Z$-periodic graphs.\myenddefinitionsymbol
\end{exmp}

In the literature, $\Z$-periodic graphs have been studied under the name of periodic (or dynamic) graphs, see, e.g., \cite{orlin1984some,hoefting1995minimum}.
In contrast, $\N$-periodic graphs do not seem to have attracted much attention.

A path $p$ in either a static or an $\S$-periodic graph is an alternating sequence $p = (v_1,e_1,v_2,\ldots,v_m)$ of nodes $v_i$ and arcs $e_i$ such that $\up(e_i) = v_i$ and $\down(e_i) = v_{i+1}$ for all $i\in\dint{1,m-1}$.
The length of $p$ is $\len(p) = m-1$.
A path $p$ is called circuit if $v_1\eqqcolon\up(p)$ and $v_m\eqqcolon\down(p)$ coincide\rerewrite{, and is called elementary if all vertices, except the first and the last one, are pairwise distinct}{}.
A path $p$ in an $\S$-periodic graph is a pseudo-circuit if $\height(v_1) = \height(v_m)$\rerewrite{, and is called pseudo-elementary if all vertices, except the first and the last one, have pairwise distinct bases}{}.
Each node (resp., arc, path, pseudo-circuit) of an $\S$-periodic graph $\graph_\S$ corresponds to a unique node (resp., arc, path, circuit) of the associated static graph $\graph$.
On the other hand, each node (resp., arc, path, circuit) of $\graph$ induces infinitely many nodes (resp., arcs, paths, pseudo-circuits) of $\graph_\N$ and $\graph_\Z$.
\rewrite{The shift, weight, and left-shift of path $p$ are defined by\footnote{In the definition of $\lshift$, we assume by convention that the empty sum is equal to $0$.} 
\[
    \shift(p) = \sum_{i=1}^{m-1} \shift(e_i),\quad
    \weight(p) = \sum_{i=1}^{m-1} w(e_i),
\]
\[
    \lshift(p) = \min_{i\in\dint{0,m-1}} \sum_{j=1}^{i} \shift(e_j).
\]}{The shift and weight of path $p$ are defined by
\[
    \shift(p) = \sum_{i=1}^{m-1} \shift(e_i),\quad
    \weight(p) = \sum_{i=1}^{m-1} w(e_i).
\]}{}
\rerewrite{}{The shift and weight of path $p$ are defined by
    $\shift(p) = \sum_{i=1}^{m-1} \shift(e_i)$,
    $\weight(p) = \sum_{i=1}^{m-1} w(e_i)$.}

\rerewrite{Note that elementary circuits in $\graph$ and pseudo-elementary pseudo-circuit in $\graph_\S$ contain at most $n$ nodes and their shift is a number in $\dint{-n,n}$.
\rewrite{Furthermore, if $p$ is a path in $\graph_\Z$, then it is also a path in $\graph_\N$ if and only if
\begin{equation}\label{eq:path_condition}
    \shift(\up(p))+\lshift(p)\in\N.
\end{equation}}{}
Let $p_1,p_2,\ldots,p_m$ be paths such that $\down(p_i) = \up(p_{i+1})$ for all $i\in\dint{1,m-1}$.
Then we write $p_1p_2\cdots p_m$ to indicate the path obtained by concatenating $p_1,p_2,\ldots,p_m$.
\rewrite{For a path $p=p_1p_2\cdots p_m$ the following formula holds:
\begin{equation}\label{eq:aux}
        \lshift(p) = \min_{i\in\dint{1,m}} \left\{\left(\sum_{j=1}^{m-1}\shift(p_j)\right) + \lshift(p_i)\right\}.
\end{equation}}{}
If $p$ is a circuit and $x\in\N$, we define $p^x = p p^{x-1}$, where $p^0$ indicates the empty path, which has zero length.}{}

%Let $v_j = (i_j,k_j)$ denote a node of an $\S$-periodic graph for all $j\in\dint{1,m}$.
%A path $p$ of length $m-1$ from $v_1$ to $v_m$ is a sequence of nodes $(v_1,v_2,\ldots,v_m)$ such that $(v_j,v_{j+1})\in E_\S$ for all $j\in\dint{1,m-1}$.
%Path $p$ is a pseudo-circuit if $i_1 = i_m$, and a circuit if also $k_1 = k_m$ (i.e., $v_{m} = v_1$).
%A pseudo-circuit $p$ is said to be pseudo-elementary (resp., elementary) if all nodes $v_h=(i_h,k_h)$, $v_j=(i_j,k_j)$ in $p$ with $h\neq j$ and $(h,j)\notin\{(1,m),(m,1)\}$ are such that $i_h\neq i_j$ (resp., $v_h \neq v_j$).

We say that an $\S$-periodic graph contains an $\infty$-weight path if there exist two nodes $u,v\in V_\S$, and an infinite sequence $p_1,p_2,\ldots$ of paths from $u$ to $v$ with increasing weight, i.e., $\weight(p_1)< \weight(p_2)< \ldots$
Consider the following decision problem.

\noindent
$\infty$-WEIGHT $\S$-PATH\\
\textbf{Instance:} Matrices $M_{-1},M_0,M_{+1}\in\Qmax^{n\times n}$.\\
\textbf{Question:} Does $\graph_\S(M_{-1},M_0,M_{+1})$ contain an $\infty$-weight path?

\begin{exmp}\label{ex:simple_graph_2}
As an illustrative example, take the graphs of Figure~\ref{fi:periodic_graph}.
For values $\alpha=-1,\beta=2$, both $\graph_\N$ and $\graph_\Z$ contain $\infty$-weight paths, as there exists a circuit with positive weight with source node $(1,k)$, for all $k$. %, e.g., $((1,1),((1,1),(1,2)),(1,2),((1,2),(1,1)))$.
Unlike finite graphs, however, $\S$-periodic graphs may contain $\infty$-weight paths even when there are no positive-weight circuits.
For example, this is the case for $\graph_\N$ and $\graph_\Z$ when $\alpha=-5,\beta=4$, as both of them contain an $\infty$-weight path from node $(1,k)$ to node $(2,k)$, for all $k$.
On the other hand, when $\alpha=-1,\beta=1$, only $\graph_\Z$ contains $\infty$-weight paths, each corresponding to a sequence of paths $p^k_1,p^k_2,\ldots$ with increasing weight from node $(1,k)$ to node $(2,k)$.
The same sequence cannot be built in $\graph_\N$, since for all $k\in\N$ and for $h$ large enough, $p^k_h$ \rewrite{does not satisfy~\eqref{eq:path_condition}, and thus }{}{}is not a path in $\graph_\N$.\myenddefinitionsymbol
\end{exmp}

\rerewrite{
In \cite[Thorem 4.8]{hoefting1995minimum}, the authors present an algorithm of polynomial time complexity to solve $\infty$-WEIGHT $\Z$-PATH.
In the present paper, we extend this result to the problem $\infty$-WEIGHT $\N$-PATH.
We will see that a slightly modified version of the algorithm by~\citeauthor{hoefting1995minimum} can be applied to solve the problem on $\N$-periodic graphs.
The proof of this fact is, however, not immediate, and requires further tools from graph theory, well partial orders, and homogeneous Diophantine equations.
\rewrite{}{For reasons of space constraints, the most technical result is stated as follows without proof.
The interested reader is invited to consult~\zor{CITE ARXIV PAPER} for the details.}{}
}{
In \cite[Thorem 4.8]{hoefting1995minimum}, the authors present an algorithm of polynomial time complexity to solve $\infty$-WEIGHT $\Z$-PATH.
This result is extended to the problem $\infty$-WEIGHT $\N$-PATH in~\zor{CITE ARXIV PAPER}, where it is shown that a slightly modified version of the algorithm by~\citeauthor{hoefting1995minimum} can be applied to solve the problem on $\N$-periodic graphs.
%The proof of this fact is, however, not immediate, and requires further tools from graph theory, well partial orders and homogeneous Diophantine equations.
}

\rewrite{\subsection{Path decompositions and compositions in static graphs}

To start, we recall from~\cite{hoefting1995minimum} the following path-decomposition technique.
Let $p = p_1 q p_2$ be a path in a static graph, where $p_1,p_2$ are (possibly empty) paths and $q = (i_j,\ldots,i_{j+l+1})$ is an elementary circuit (thus $i_j = i_{j+l+1}$) such that all inner nodes $i_{j+1},\ldots,i_{j+l}$ of $q$ are contained somewhere else in $p$ (i.e., in $p_1$ or in $p_2$).
Since $q$ is a circuit, we can "cut it out" from $p$; the remaining path $p_1p_2$ has $l$ nodes and $l$ arcs less than $p$.
We repeat this circuit elimination until such circuits do not exist anywhere in the remaining path.
If through this procedure we have removed $x_j$ elementary circuits $q_j$, and the remaining path is $p'$, then we say that $\pazocal{C}_p = (p',\{(q_1,x_1),\ldots,(q_m,x_m)\})$ is a \emph{complete path decomposition} (CPD) of $p$.
The remaining path $p'$ constructed in this way has always less than $r^2$ nodes, where $r\leq n$ is the number of different nodes in $p$ (see~\cite{hoefting1995minimum}).

%Let $\hat{p}$ be a path in $\graph_\N$, and let $p$ be the corresponding path in $\graph$ with complete path decomposition $(p',(q_1,x_1,i_1,s_1,w_1),\ldots,(q_m,x_m,i_m,s_m,w_m))$.
%Assume that in the complete path decomposition there are no circuits $q_j$ with $s_j = 0$ and no $q_j\neq q_h$ with $(i_j,s_j) = (i_h,s_h)$.
%We now show that, under these conditions, we can always construct another path $\hat{r}$ in $\graph_\N$ % with $\shift(\up(\hat{r})) = \shift(\up(\hat{p}))+n$,
%with corresponding path in $\graph$ $r$ obtained from $p$ by permuting its subpaths such that their order does not depend on $x_j$. % all elementary circuits $q_j$ with positive shift and postponing those with negative shift.
We now introduce a "canonical" method to reconstruct a path from a CPD, such that the resulting path has some convenient features.
Let $p$ be a path in the static graph $\graph$ with CPD $\pazocal{C}_p = (p',\{(q_1,x_1),\ldots,(q_m,x_m)\})$.
Then, we can factor $p$ as $p = r_1r_2\ldots r_l$, where each $r_i$ is either a path with $\len(r_i) = 1$ or one of the $x_j$ occurrences of elementary circuit $q_j$ that are "cut out" from $p$ through the CPD to obtain $p'$.
Suppose that $\pazocal{C}_p$ satisfies the following condition.

\noindent
\textit{Assumption A}: all elementary circuits $q_j$ from $\pazocal{C}_p$ have nonzero shift, and for all $q_j\neq q_h$, $(\up(q_j),\shift(q_j))\neq (\up(q_h),\shift(q_h))$.\myenddefinitionsymbol

Then, we can construct another path $t$ with the same CPD $\pazocal{C}_p$ as follows:

\renewcommand{\labelenumi}{\theenumi.}
\renewcommand{\theenumi}{\arabic{enumi}}
\begin{enumerate}
    \item initialize $s\leftarrow -n$, $i\leftarrow 1$, $r\leftarrow p=r_1r_2\ldots r_l$, and label from 1 to $x_j$ each occurrence of each elementary circuit $q_j$ in $p$ that is "cut out" from $p$ through the CPD;
    \item\label{en:q_j} let $q_j$ be the circuit from $\pazocal{C}_p$ with $\shift(q_j)=s$ and $\up(q_j)=i$.
    If no such circuit exists, go to Step~\ref{en:s_1};
    \item\label{en:swap} perform the operations described in the remainder of this step for every $\xi\in\dint{1,x_j}$.
    Let $r_h$ be the occurrence of $q_j$ labeled $\xi$ in $r$.
    Define
    \begin{equation}\label{eq:rplus}
            r^+ = 
            \begin{dcases}
                r_1\cdots r_{h-1}r_{h+1} \cdots r_y r_hr_{y+1} \cdots r_l & \mbox{if } s<0,\\
                r_1\cdots r_{y}r_{h}r_{y+1} \cdots r_{h-1} r_{h+1} \cdots r_l & \mbox{if } s>0,
            \end{dcases}
    \end{equation}
    where $y$ is chosen such that, if $s<0$ it is the maximal, and if $s>0$ it is the minimal number in $\dint{0,l}$ such that $\up(r_h)=\down(r_y)$ or $\down(r_h)=\up(r_{y+1})$.
    Path $r^+$ could thus be written as $r^+ = r_{\sigma^+(1)} r_{\sigma^+(2)} \cdots r_{\sigma^+(l)}$, where $\sigma^+$ is an appropriate permutation of $\dint{1,l}$.
    Redefine $r_h\leftarrow r_{\sigma^+(h)}$ for all $h\in\dint{1,l}$, and $r\leftarrow r^+$;
    \item\label{en:s_1} set $i\leftarrow i+1$.
    If $i = n+1$, then set $s\leftarrow s+1$ and $i\leftarrow 1$. 
    If $s=n+1$, then set $t\leftarrow r$ and stop the algorithm, otherwise go to Step~\ref{en:q_j}.
\end{enumerate}
In words, the above algorithm constructs $t$ from $p$ by postponing elementary circuits with negative shift and anticipating elementary circuits with positive shift as much as possible.
The output is a path with factorization $t = t_1 q_{\sigma(1)}^{x_{\sigma(1)}} t_2 q_{\sigma(2)}^{x_{\sigma(2)}}\cdots q_{\sigma(m)}^{x_{\sigma(m)}} t_{m+1}$, where $t_1t_2\cdots t_{m+1} = p'$ and $\sigma$ is a permutation of $\dint{1,m}$.
Note that, if $p_1$ and $p_2$ are two paths with the same CPD $\pazocal{C}_p$, then applying the above algorithm to $p_1$ and $p_2$ would generate the same output $t$.
Therefore, if $t$ is the path obtained from the algorithm above, then we say that it is the \emph{canonical path composition} (CPC) of the CPD $\pazocal{C}_p$.

The CPC $t$ of a CPD $\pazocal{C}_p$ has two useful properties.
The first one is that both the factorization of $p'$ into $t_1t_2\cdots t_{m+1}$ and the permutation $\sigma$ are independent of the number of repetitions $x_j$ of each elementary circuit $q_j$ in $\pazocal{C}_p$.
Therefore, given two CPDs, $\pazocal{C}_{p_1}$ and $\pazocal{C}_{p_2}$, where $\pazocal{C}_{p_h}=(p',\{(q_1,x_{h,1}),\ldots,(q_m,x_{h,m})\})$ for $h\in\{1,2\}$, differing only in the number of repetitions of elementary circuits, their respective CPCs, $t^{(1)}$ and $t^{(2)}$, are of the form $t^{(h)} = t_1 q_{\sigma(1)}^{x_{h,\sigma(1)}} t_2 q_{\sigma(2)}^{x_{h,\sigma(2)}}\cdots q_{\sigma(m)}^{x_{h,\sigma(m)}} t_{m+1}$.
%The second property of a CPC is that $t$ can always be obtained from $p$ by "anticipating" elementary circuits with positive shift and "postponing" elementary circuits with negative shift.
The second property of a CPC is given by the following lemma. 

\begin{lem}\label{le:CPCproperty}
%Let $\hat{p}$ be a path in the $\N$-periodic graph $\graph_\N$ with $\up(\hat{p}) = (i,k)$, and let $p$ be the corresponding path in the static graph $\graph$ with CPD $\pazocal{C}_p$.
Let $p$ be a path in the static graph $\graph$, such that the CPD $\pazocal{C}_p$ satisfies Assumption A.
Let $t$ be the CPC of $\pazocal{C}_p$, and $n$ be the number of nodes of $\graph$.
Then $\lshift(t) \geq \lshift(p) - n$.\myendtheoremsymbol
%Then the unique path $\hat{t}$ in $\graph_\Z$ induced by $t$ and such that $\up(\hat{t}) = (i,k+n)$ is also a valid path in $\graph_\N$.
\end{lem}
\begin{pf}
%To prove that $\hat{t}$ is a valid path of $G_\N$ we need to verify that it satisfies~\eqref{eq:path_condition}.
%Clearly, if $p$ is the CPC of $\pazocal{C}_p$, then $t$ and $p$ coincide, and $\hat{t}$ is obtained from $\hat{p}$ by shifting its source node by $+n$; therefore, it is a valid path in $\graph_\N$.
%
%Suppose instead that $p$ and $t$ do not coincide.
The CPC $t$ of $\pazocal{C}_p$ is obtained from $p$ by anticipating elementary circuits with positive shift and postponing elementary circuits with negative shift.
We factor a generic path $r$ generated in Step~\ref{en:swap} of the algorithm as $r = r_1r_2\cdots r_l$, where each $r_h$ is either a path of unitary length or an elementary circuit whose position is shifted during the execution of the algorithm.
We now show by induction that every path $r$ obtained during the execution of the algorithm (therefore, also $t$) satisfies the following property: for all $j\in\dint{1,l}$, $\sum_{d=1}^{j}\shift(r_d)\geq \lshift(p)$.
Since $\lshift(r_j)\geq -n$ for all $j$, if this property holds then, from~\eqref{eq:aux}, we have
\[
    \begin{array}{rl}
    \lshift(r) &=
    \begin{aligned}[t] \min_{j\in\dint{1,l}} \left\{\left(\sum_{d=1}^{j-1}\shift(r_d)\right) + \lshift(r_j)\right\}
    \end{aligned}\\
        & \geq \lshift(p) - n,
    \end{array}
\]
which implies that $\lshift(t) \geq \lshift(p)-n$.
%\begin{align*}
%    & \lshift(r) =\\
%    & = \begin{aligned}[t] \min_{j\in\dint{1,l}} \left\{\left(\sum_{d=1}^{j-1}\shift(r_d)\right) + \lshift(r_j)\right\}
%    \end{aligned}\\
%        & \geq \lshift(p) - n.
%\end{align*}
%which, combined with $\shift(\up(\hat{t}))+\lshift(\hat{t})=k+n+\lshift(t)$, would immediately imply that $\hat{t}$ satisfies~\eqref{eq:path_condition}.

The base case corresponds to $r=p$, for which the property $\sum_{d=1}^j \lshift(r_d) \geq \lshift(p) \geq \lshift(p)-n$ obviously holds for all $j$.
%By means of contradiction, suppose that $k+\sum_{d=1}^j \shift(r_d)< 1$ for some $j$.
%But then $\shift(\up(\hat{p}))+\lshift(\hat{p}) = k+ \lshift(p) < 1$, implying that $\hat{p}$ is not a path in $\graph_{\N}$, which violates one of the hypotheses of the proposition.

Suppose now that $r$ satisfies the property, and let $r^+$ be the path obtained in Step~\ref{en:swap} from $r$ by shifting one elementary circuit; if $r_h$ is the elementary circuit being shifted, with $\shift(r_h)=s$, then we can define $r^+=r_{\sigma^+(1)}r_{\sigma^+(2)}\cdots r_{\sigma^+(l)}$ as in~\eqref{eq:rplus}
%\[
%    \begin{array}{rl}
%    r^+ \!\! &= 
%    \begin{dcases}
%        r_1\cdots r_{i-1}r_{i+1} \cdots r_y r_ir_{y+1} \cdots r_l & \mbox{if } \shift(r_i)<0,\\
%        r_1\cdots r_{y}r_{i}r_{y+1} \cdots r_{i-1} r_{i+1} \cdots r_l & \mbox{if } \shift(r_i)>0
%    \end{dcases}\\
%    &= r_{\sigma(1)}\cdots r_{\sigma(l)},
%\end{array}
%\]
for some $y$ and permutation $\sigma^+$.

Consider first the case $\shift(r_h)<0$.
Then, for all $j\in \dint{1,i-1}\cup\dint{y,l}$, 
\[
    \sum_{d=1}^j \shift(r_{\sigma^+(d)}) = \sum_{d=1}^j \shift(r_d)\geq \lshift(p) 
\] 
by the induction hypothesis.
Moreover, for all $j\in\dint{i,y-1}$, 
\[
    \begin{array}{rcl}
    \displaystyle\sum_{d=1}^j \shift(r_{\sigma^+(d)}) &=& \displaystyle\sum_{d=1}^{i-1} \shift(r_{d})+\sum_{d=i+1}^{j} \shift(r_{d}) \\
    &>& \displaystyle\sum_{d=1}^{j} \shift(r_d)\geq \lshift(p)
    \end{array}
\]
by the fact that $\shift(r_h)<0$ and the induction hypothesis.

Now consider the case $\shift(r_h)>0$.
For all $j\in\dint{1,y}\cup\dint{i,l}$, 
\[
    \sum_{d=1}^j \shift(r_{\sigma^+(d)}) = \sum_{d=1}^j \shift(r_d)\geq \lshift(p)
\]
by the induction hypothesis.
Finally, for all $j\in\dint{y+1,i-1}$, 
\[
    \begin{array}{rcl}
    \displaystyle\sum_{d=1}^j \shift(r_{\sigma^+(d)}) &=& \displaystyle\sum_{d=1}^{j-1} \shift(r_d) + \shift(r_h) \\
    &>&\displaystyle \sum_{d=1}^{j-1} \shift(r_d) \geq \lshift(p)
    \end{array}
\]
by the fact that $\shift(r_h)>0$ and the induction hypothesis. \myendproofsymbol
\end{pf}

\subsection{Well partial orders and Diophantine equations}

Before providing the algorithm to solve $\infty$-WEIGHT $\N$-PATH, in this section we recall some concepts from well partial orders and homogeneous Diophantine equations.

Here is a possible definition of well partial order (see, e.g., \cite{KRUSKAL1972297,higman1952ordering}).
A \textit{well partial order} (wpo) $\preceq$ over a set $X$ is a partial order relation such that, for any infinite sequence $x_1,x_2,x_3,\ldots$ of elements from $X$, there is an infinite nondecreasing subsequence $x_{i_1}\preceq x_{i_2}\preceq x_{i_3}\preceq \ldots$, with $i_1<i_2<i_3<\ldots$
Examples of wpos are $\leq$ over $\N$, and its component-wise extension over $\N^n$, defined for all $a,b\in\N^n$ by: $a \leq b$ if and only if $a_i \leq b_i$ for all $i\in\dint{1,n}$.
A useful property of wpos is that, if $\preceq$ is a wpo over $X$, then it is a wpo over any subset $Y$ of $X$.
Another example of wpo comes from the following lemma.

\begin{lem}\label{le:wpoCPD}
    Let $\pazocal{C}_p = (p',\{(x_1,q_1),\ldots,(x_l,q_l)\})$ and $\pazocal{C}_t = (t',\{(y_1,r_1),\ldots,(y_m,r_m)\})$ be two CPDs of a given static graph $\graph$.
    The partial order relation $\preceq_{\textup{CPD}}$ defined over the set of CPDs of $\graph$ by $\pazocal{C}_p \preceq_{\textup{CPD}} \pazocal{C}_t$ if and only if $p'=t'$, $l=m$, $(q_1,\ldots,q_l)=(r_1,\ldots,r_l)$, and 
    \[
        \begin{bmatrix}x_1\\ x_2\\ \vdots\\ x_l\end{bmatrix}\leq \begin{bmatrix}y_1\\ y_2\\ \vdots\\ y_l\end{bmatrix} 
    \]
    is a wpo.\myendtheoremsymbol
\end{lem}
\begin{pf}
    Observe that, for each static graph $\graph$, the number of paths $p'$ and elementary circuits $q_h$ in all possible CPDs $(p',\{(x_1,q_1),\ldots,(x_l,q_l)\})$ is finite, since the number of nodes $n$ of $\graph$ is finite and the lengths of $p'$ and $q_h$ are bounded by a polynomial function in $n$.
    Therefore, given an infinite sequence of CPDs, there is an infinite subsequence with the same $p'$, $l$, and $(q_1,\ldots,q_l)$.
    To conclude, recall that $\leq$ is a wpo over $\N^l$ and that $x_h\in\N$ for all $h\in\dint{1,l}$. \myendproofsymbol
\end{pf}

Let $A$ be an $m\times n$ matrix with coefficients from $\Z$ and let $\vec{0}$ indicate a vector (whose dimension will be made clear from the context) whose entries are all $0$.
A \textit{homogeneous linear Diophantine system} is an equation of the form $Ax = \vec{0}$, where $x\in\N_0^{n}$ is the vector of variables.
The solution set of $Ax =\vec{0}$ is $S_A = \{x\in\N_0^n\mid Ax = 0\}$. %, and say that $x$ is a minimal solution if $x\in S_A\setminus\{\vec{0}\}$ and, for all $y\in S_A\setminus\{\vec{0}\}$, $y \not< x$.
The carrier (or support) $\pazocal{C}(x)$ of vector $x$ is the set of indices of the nonzero elements in $x$, i.e., $\pazocal{C}(x) = \{i\mid x_i\neq 0\}$.
%Then, we say that a solution $x\in S_A\setminus\{\vec{0}\}$ is of minimal carrier if for all $y\in S_A\setminus\{\vec{0}\}$, $\pazocal{C}(y) \not\subset \pazocal{C}(x)$.
We define the set $S^{\min}_A$ of minimal solutions of $Ax = \vec{0}$ with minimal carrier by
\[
    \begin{array}{rl}
    S^{\min}_A = \{x\in S_A\setminus \{\vec{0}\}\mid & \forall y\in S_A\setminus\{x,\vec{0}\}, \\
    & (y \not\leq x \wedge \pazocal{C}(y) \not\subset \pazocal{C}(x))\}.
    \end{array}
\]
Since $y\neq x$, the condition $x\not\leq y$ means that there exists $i\in\dint{1,n}$ such that $y_i > x_i$ (i.e., $x$ is a minimal solution), and $\pazocal{C}(y)\not\subset \pazocal{C}(x)$ indicates that either $\pazocal{C}(y)=\pazocal{C}(x)$ or there exists $j\in\dint{1,n}$ such that $y_j>0$ and $x_j=0$ (i.e., $x$ has minimal carrier).
Given a set $X\subseteq \R^n$, the conical hull of $X$ is
\[
    \mbox{Cone}(X) \coloneqq \left\{\sum_{i=1}^l \lambda_i x_i\mid l\in\N_0,\ x_i\in X,\ \lambda_i\in\R,\ \lambda\geq 0\right\}.
\] 
\begin{thm}\label{th:diophantine}
    \cite[Theorem 3]{domenjoud1991solving}
    Every solution $x\in \N_0^n$ of $Ax = \vec{0}$ is a linear combination with nonnegative coefficients of vectors from $S^{\min}_A$, i.e.,
    \[
        S_A \subseteq \mbox{Cone}(S^{\min}_A) .\tag*{\myendtheoremsymbol}
    \]
\end{thm}

\subsection{The algorithm}

This section provides the polynomial-time algorithm to solve $\infty$-WEIGHT $\N$-PATH and the proof of its correctness.
}
{}{}
\rerewrite{
\begin{lem}\label{le:main}
    Let $G^\N$ be an $\N$-periodic graph induced by a static graph $G$ with $n$ nodes.
    Then $G^\N$ contains an $\infty$-weight path if and only if $G$ contains either a circuit with length at most $n$, positive weight, and zero shift, or a path $p = q_1 p' q_2$, where $\len(p')<n$, $\len(q_1),\len(q_2)\leq n$, and $q_1,q_2$ are circuits with shifts $s_1,s_2$ and weights $w_1,w_2$, respectively, satisfying
    \begin{equation}
        s_1 > 0,\quad s_2 < 0, \label{eq:le1}
    \end{equation}
%    \begin{equation}
%        x_1 = \frac{-s_2}{\gcd(s_1,-s_2)}, \quad  x_2 = \frac{s_1}{\gcd(s_1,-s_2)}, \label{eq:le2}
%    \end{equation}
    \begin{equation}
       -s_2 w_1 + s_1 w_2 > 0. \label{eq:le3}
    \end{equation}
\myendtheoremsymbol
\end{lem}
}{}
\rewrite{
\begin{pf}
"$\Leftarrow$": the existence of a circuit with positive weight and zero shift in $\graph$ implies the presence of a circuit with positive weight in $\graph_\N$ and, thus, of an $\infty$-weight path.

Suppose that no circuit with length at most $n$, positive weight, and zero shift exists in $\graph$, but that there is a path $p=q_1 p' q_2$ satisfying the properties stated in the theorem.
Let us construct another path $r = q_1^{x_1} p' q_2^{x_2}$, where $x_1 = -s_2$ and $x_2 = s_1$.
Clearly,
\begin{equation}\label{eq:le22}
    x_1 s_1 + x_2 s_2 = 0.
\end{equation}
Through some simple calculations involving~\eqref{eq:aux}, \eqref{eq:le1}, \eqref{eq:le22}, and the fact that $q_1$ and $q_2$ have length at most $n$, it is possible to compute the following upper bound for $\lshift(r)$: $\lshift(r)\geq b = \lshift(p')-n$. % \min\{a-n,2a-n,3a\}$, where $a = \min_{j\in\dint{1,3}} (\lshift(p_j))$.
Therefore, the path $\hat{r}$ induced by $r$ and with $\up(\hat{r})= (\up(r),1-b)$ and $\weight(\hat{r}) = \weight(r)$ is a valid path of $G_\N$, as it satisfies~\eqref{eq:path_condition}.
Since the number $b$ does not depend on $x_1,x_2$, all paths of the form $r^{(h)} = q_1^{hx_1}p' q_2^{hx_2}$, where $h\in\N$, satisfy $\lshift(r^{(h)})\geq b$.
Thus, $r^{(1)},r^{(2)},\ldots$ induce in $G_\N$ paths $\hat{r}^{(1)},\hat{r}^{(2)},\ldots$ with $\up(\hat{r}^{(h)}) = (\up(r),1-b)$ and, because of~\eqref{eq:le3}, $\weight(\hat{r}^{(1)})<\weight(\hat{r}^{(2)})<\ldots$
This implies that $G_\N$ contains an $\infty$-weight path.

"$\Rightarrow$": let $\hat{p}^{(1)},\hat{p}^{(2)},\ldots$ be the infinite sequence of paths in $\graph_\N$ from node $(i,k)$ to node $(i',k')$ with increasing weight, and let $p^{(1)},p^{(2)},\ldots$ be the infinite sequence of paths $p^{(h)}$ in $\graph$ from node $i$ to node $i'$ corresponding to $\hat{p}^{(h)}$.
In the remainder of this proof, we will use the notation $\shift(q_j) \eqqcolon s_j$ and $\weight(q_j) \eqqcolon w_j$ for any elementary circuit $q_j$ of the CPD of any path $p^{(h)}$.

From Lemma~\ref{le:wpoCPD}, there is an infinite subsequence of paths $p^{(i_1)},p^{(i_2)},\ldots$ in the static graph with nondecreasing CPDs (in the sense of $\preceq_{\textup{CPD}}$) and increasing weight.
Without loss of generality, we can suppose that Assumption A holds for the CPD of each of these paths.
Indeed, if there is an elementary circuit $q_h$ with $s_h = 0$ in the CPD of one of the paths, then either $w_h \leq 0$, in which case we can remove it obtaining a path with larger or equal weight and left-width, or $w_h>0$, and therefore $q_h$ is a circuit with length less than $n$, zero shift, and positive weight.
On the other hand, suppose that there are two elementary circuits $q_j\neq q_h$ with equal shift and source in the CPD of one of the paths $p^{(h)}$, with $w_j\leq w_h$.
Then, substituting each occurrence of $q_j$ with an occurrence of $q_h$ (and repeating this procedure for any pair of circuits with equal shift and source) results in a path with larger weight and with left-shift lower by at most $n$ (from~\eqref{eq:path_condition} and the fact that $q_j$ and $q_h$ are elementary).
%To make the corresponding path, say $\bar{\hat{p}}^{(h)}$, valid in $\graph_\N$, it is then sufficient to define its source node by $(i,k+n)$.

Since Assumption A holds, from the infinite sequence of paths $p^{(i_1)},p^{(i_2)},\ldots$ we can construct another sequence $t^{(1)},t^{(2)},\ldots$ such that $t^{(h)}$ is the CPC of $\pazocal{C}_{p^{(i_h)}}$.
As $\lshift(p^{(i_h)})\geq -k$ (otherwise $\hat{p}^{(i_h)}$ would not be a valid path in $\graph_\N$), from Lemma~\ref{le:CPCproperty} we have that $\lshift(t^{(h)})\geq -k -n$ for all $h$.
Recall that each $t^{(h)}$ can be factored into $t^{(h)} = t_1 q_1^{x_{1,h}} t_2 q_2^{x_{2,h}}\cdots q_m^{x_{m,h}} t_{m+1}$.
Then, because of~\eqref{eq:aux} and the bound on $\lshift(t^{(h)})$, we can construct the bound $b = k+n+\max_{l\in\dint{1,m}} \shift(t_1\cdots t_l)$, for which the following expression holds true for all $h\in\N$ and $l\in\dint{1,m}$:
\begin{equation}\label{eq:shiftbound}
    \sum_{j=1}^{l} s_j x_{j,h} + b \in \No.
\end{equation}
Since~\eqref{eq:shiftbound} must hold for all $h\in\N$, and since $\leq$ is a wpo over $\No$, we can find a subsequence of $t^{(1)},t^{(2)},\ldots$, say $t^{(i_1)},t^{(i_2)},\ldots$ for which
\begin{equation}\label{eq:auxx}
    \sum_{j=1}^{l} s_jx_{j,i_{h+1}} + b \geq \sum_{j=1}^{l} s_jx_{j,i_h} + b.
\end{equation}
Take any $h\in\N$, and define, for all $j\in\dint{1,m}$, $y_j \coloneqq x_{j,i_{h+1}} - x_{j,i_h}$.
Note that, for all $j$, $y_{j}\in\No$ because $\pazocal{C}_{t^{(i_{h})}} \preceq_{\textup{CPD}} \pazocal{C}_{t^{(i_{h+1})}}$.
For the selected $h$, \eqref{eq:auxx} can then be rewritten as
\begin{equation}\label{eq:final1}
    \sum_{j=1}^{l} s_jy_{j} \geq 0 \quad \quad \forall l\in\dint{1,m-1}.
\end{equation}
%where $y_{j,h} = x_{j,i_{h+1}} - x_{j,i_h}$ for all $l\in\dint{1,m}$.
Additionally, as $t^{(i_h)}$ and $t^{(i_{h+1})}$ have the same shift,
\[
    \sum_{j=1}^{m} s_jx_{j,i_{h+1}} + \shift(t') = \sum_{j=1}^{m} s_jx_{j,i_h} + \shift(t'),
\]
where $t' = t_1t_2\cdots t_m$.
This implies the following, stronger condition for $l=m$:
\begin{equation}\label{eq:final2}
    \sum_{j=1}^{m} s_jy_{j} = 0.
\end{equation}
Since $\weight(t^{(i_h)})<\weight(t^{(i_{h+1})})$, we also have,
\[
   \sum_{j=1}^{m} w_jx_{j,i_{h+1}} + \weight(t') > \sum_{j=1}^{m} w_jx_{j,i_h} + \weight(t'),
\]
which is equivalent to
\begin{equation}\label{eq:final3}
    \sum_{j=1}^{m} w_jy_{j} > 0 .
\end{equation}

We now rewrite~\eqref{eq:final1} and~\eqref{eq:final2} as the homogeneous linear Diophantine system $A x = \vec{0}$, where
\[
    x = 
    \begin{bmatrix}
        y_1 & \cdots & y_m & | & z_1 & \cdots & z_{m-1}
    \end{bmatrix}^{\top} = 
    \begin{bmatrix}
        y^\top & | & z^\top
    \end{bmatrix}^{\top}
    \in\No^{2m-1},
\]
and
\newenvironment{smallarray}[1]
 {\null\,\vcenter\bgroup\scriptsize
  \renewcommand{\arraystretch}{0.7}%
  \arraycolsep=.13885em
  \hbox\bgroup$\array{@{}#1@{}}}
 {\endarray$\egroup\egroup\,\null}
\[
    A = \left[
    \begin{smallarray}{cccccc|ccccc}
        s_1 & 0 & 0 & \cdots & 0 & 0 & -1 & 0 & 0 & \cdots & 0 \\
        0 & s_2 & 0 & \cdots & 0 & 0 & 1 & -1 & 0 & \cdots & 0 \\
        0 & 0 & s_3 & \cdots & 0 & 0 & 0 & 1 & -1 & \cdots & 0 \\
        \svdots & \svdots & \svdots & \sddots & \svdots & \svdots & \svdots & \svdots & \svdots & \sddots & \svdots \\
        0 & 0 & 0 & \cdots & s_{m-1} & 0 & 0 & 0 & 0 & \cdots & -1\\ 
        0 & 0 & 0 & \cdots & 0 & s_m & 0 & 0 & 0 & \cdots & 1 
    \end{smallarray}\right]\in\Z^{m\times (2m-1)}.
\]
Note that each element $z_l$ of the variables vector $x$ is forced to take the value $\sum_{j=1}^{l} s_j y_j$ by the Diophantine system.
In the following, we construct the set $S_A^{\min}$ of minimal solutions of $Ax=\vec{0}$ with minimal carrier.
Let $x=[y^\top\ z^\top]^\top \in S_A^{\min}$.
In order to satisfy $Ax=\vec{0}$, $y=\vec{0}$ implies $z=\vec{0}$.
Then at least one element of $y$ must be nonzero.
Let $i$ be the minimal index such that $y_{i}>0$.
This forces $z_l$ to be $0$ for all $l \in\dint{1,i-1}$ and $z_i = s_iy_i$.
Then, in order for $z_{i}$ to be nonnegative, we must have $s_{i}>0$.
Moreover, to achieve~\eqref{eq:final2}, there must exist a second element $y_j>0$ corresponding to a shift $s_j<0$.
Let $j$ be the minimal index (greater than $i$) for which $y_j>0$ and $s_j<0$.
Due to the minimality of $i$ and $j$, $z_l$ must be positive and equal to $s_iy_i$ for all $l\in\dint{i+1,j-1}$, and $z_j = s_i y_i + s_j y_j$.
To achieve minimal carrier, we impose $s_i y_i + s_j y_j = 0$; observe that this Diophantine equation admits minimal solution $y_i = s_{ij}$ and $y_j = s_{ji}$, where $s_{hk} = \frac{|s_k|}{\gcd(|s_h|,|s_k|)}$.
Therefore, from the minimality of $x$ and of its carrier, it is necessary to have $y_i = s_{ij}$, $y_j = s_{ji}$, $y_l = 0$ for all $l\notin\{i,j\}$, and $z_l = 0$ for all $l\in\dint{j,m-1}$.
This reasoning shows that $S_A^{\min}$ coincides with the set of elements $x=[y^\top\ z^\top]^\top$ such that
\[
    y_l=
    \begin{dcases}
        s_{ij} & \mbox{if } l=i,\\
        s_{ji} & \mbox{if } l=j,\\
        0 & \mbox{otherwise},
    \end{dcases}\quad
    z_l = \begin{dcases}
        s_i s_{ij} & \mbox{if } l\in\dint{i,j-1},\\
        0 & \mbox{otherwise},
    \end{dcases}
\]
where $i<j$, $s_i>0$, and $s_j<0$.

From Theorem~\ref{th:diophantine}, if $y_1,\ldots,y_m$ solve~(\ref{eq:final1}-\ref{eq:final3}), then $y = [y_1\cdots y_m]^\top$ can be written as
\[
    y = \sum_{\{i,j\mid s_i>0, s_j<0\}} \lambda_{ij} y_{ij},\quad 
    (y_{ij})_l =
    \begin{dcases}
        s_{ij} & \mbox{if } l = i,\\
        s_{ji} & \mbox{if } l = j,\\
        0 & \mbox{otherwise,}
    \end{dcases}
\]
where $\lambda_{ij}\geq 0$ for all $i,j$.
By substituting this expression of $y$ into~\eqref{eq:final3} we get
\[
    \left(\sum_{\{i,j\mid s_i>0, s_j<0\}} \lambda_{ij} y_{ij}\right)\cdot w = \sum_{\{i,j\mid s_i>0, s_j<0\}} \lambda_{ij} (y_{ij} \cdot w) >0,
\]
where $w = [w_1 \cdots w_m]^\top$ and $\cdot$ is the dot product.
But this inequality implies that, for at least one pair $(i,j)$, 
\[
    y_{ij}\cdot w = y_{i}w_{i} + y_{j}w_{j} = \frac{-s_jw_i+s_iw_j}{\gcd(s_i,-s_j)}>0,
\]
where $s_{i}>0$ and $s_{j}<0$.
Hence, $-s_j w_i + s_i w_j >0$.

We can now define path $t_1 q_1^{(y_{ij})_1} t_2 q_2^{(y_{ij})_2} \cdots q_m^{(y_{ij})_m} t_{m+1}$, in which only $(y_{ij})_i$ and $(y_{ij})_j$ are nonzero.
By renaming $q_1 \coloneqq q_{i}$, $q_2 \coloneqq q_j$, $\bar{p} \coloneqq t_{i+1} \cdots t_{j}$, we have that path $p = q_1 \bar{p} q_2$ satisfies the statement of the lemma except, possibly, for $\len(\bar{p})<n$ (recall indeed that every elementary circuit of $\graph$ has length at most $n$).
Finally, observe that, since $\graph$ has $n$ nodes, if there is a path $\bar{p}$ in $\graph$ from $\down(q_1)$ to $\up(q_2)$, then there is also an (elementary) path $p'$ from $\down(q_1)$ to $\up(q_2)$ with $\len(p')<n$.\myendproofsymbol
\end{pf}
}{}{}

\rerewrite{
Because of Lemma~\ref{le:main}, we can \rewrite{finally}{}{} state the main theorem of this section.
Recall that, using the Floyd-Warshall algorithm, it is possible to verify the existence of circuits with positive weight in a graph with $n$ nodes, and, if no such circuits exist, to verify the existence and compute the largest weight of paths between all pairs of nodes\footnote{From an algebraic perspective, the Floyd-Warshall returns the max-plus Kleene star of the adjacency matrix corresponding to the given graph. Such concepts are defined in Section~\ref{se:algebra}.} in time $\pazocal{O}(n^3)$ (see, e.g., \cite{cormen2022introduction}).
}{}

\begin{thm}\label{th:polynomial_complexity}
The problem $\infty$-WEIGHT $\N$-PATH is solvable in strongly polynomial time complexity $\pazocal{O}(n^6)$.\rerewrite{\footnote{It is possible to reduce the time complexity of the algorithm at least to $\pazocal{O}(n^5)$ by exploiting the sparsity of graph $\graph_{\dint{-n,n}}$, defined in the proof of Theorem~\ref{th:polynomial_complexity}. Proving this is out of the scope of the present paper.}}{}\myendtheoremsymbol
\end{thm}
\rerewrite{
\begin{pf}
Given matrices $M_{-1},M_0,M_{+1}\in\Qmax^{n\times n}$, we need to decide whether $\graph_\N = \graph_{\N}(M_{-1},M_0,M_{+1})$ contains an $\infty$-weight path.
From Lemma~\ref{le:main}, this is equivalent to decide whether the static graph $\graph = \graph(M_{-1},M_0,M_{+1})$ contains either a circuit with length at most $n$, zero shift, and positive weight, or a path $p = q_1p'q_2$, where $q_1,q_2$ are circuits with $\len(q_1),\len(q_2)\leq n$, $\len(p')<n$, and \eqref{eq:le1}-\eqref{eq:le3} hold.
Observe that every circuit $q$ in $\graph$ with length $\leq n$ induces a pseudo-circuit $\hat{q}$ in $\graph_{\dint{-n,n}}$ with $\up(\hat{q}) = (\up(q),0)$.
%Thus, we can search such pseudo-circuits in $\graph_{\dint{-n,n}}$ instead of $\graph$.
Thus, we can focus on $\graph_{\dint{-n,n}}$ instead of $\graph$.
In the following, the polynomial-time algorithm is stated.
%Thus, we can use the following algorithm to search the path $p$ in $\graph$.

%First, check if graph $\graph_{\dint{-n,n}}$ contains circuits with positive weight; .
In time complexity $\pazocal{O}(n^3)$, use the Floyd-Warshall algorithm to compute matrix $H^*\in\{0,1\}^{n\times n}$, whose entry $(i,j)$ is $1$ if there is a path $p'$ in $\graph$ from node $j$ to node $i$ and $0$ otherwise.
Apply the Floyd-Warshall algorithm on graph $\graph_{\dint{-n,n}}$ (which consists of $(2n+1)n$ nodes), in time $\pazocal{O}([(2n+1)n]^3) = \pazocal{O}(n^6)$, to check if it contains circuits with positive weight; if yes, then $\graph$ contains a circuit with length at most $n$, zero shift, and positive weight, and $\graph_\N$ contains an $\infty$-weight path.
If no, the same call of the Floyd-Warshall algorithm provides, for all $z\in\{1,2\}$, $i_z\in\dint{1,n}$, $s_1\in\dint{1,n}$, $s_2\in\dint{-n,-1}$, the largest weight $w(i_z,s_z)=\weight(\hat{q}_z)$ of all pseudo-circuits $\hat{q}_z$ in $\graph_{\dint{-n,n}}$ with $\up(\hat{q}_z)=(i_z,0)$ and $\down(\hat{q}_z)=(i_z,s_z)$ (if such pseudo-circuit $\hat{q}_z$ does not exist, then set $w(i_z,s_z)=-\infty$).
%This computation can be done in time $\pazocal{O}([(2n+1)n]^3) = \pazocal{O}(n^6)$ by applying the Floyd-Warshall algorithm to $G_{\dint{-n,n}}$.
For all tuples $(i_1,i_2,s_1,s_2)$, check if $H^*_{i_2,i_1}=1$, $w(i_1,s_1),w(i_2,s_2)\neq -\infty$, and
\[
    -s_2 w(i_1,s_1)+s_1 w(i_2,s_2)>0 .
\]
This step of the algorithm takes $\pazocal{O}(n^4)$ operations.
The graph $\graph_\N$ contains an $\infty$-weight path if and only if, for at least one tuple, the three conditions are satisfied. \myendproofsymbol
\end{pf}
}{}

\rerewrite{}{
The algorithm is presented in the following, together with the analysis of its time complexity; for more details, please consult~\zor{CITE ARXIV PAPER}.
Recall that, using the Floyd-Warshall algorithm, it is possible to verify the existence of circuits with positive weight in a graph with $n$ nodes, and, if no such circuits exist, to verify the existence and compute the largest weight of paths between all pairs of nodes\footnote{From an algebraic perspective, the Floyd-Warshall returns the max-plus Kleene star of the adjacency matrix corresponding to the given graph. Such concepts are defined in Section~\ref{se:algebra}.} in time $\pazocal{O}(n^3)$ (see, e.g., \cite{cormen2022introduction}).

In time complexity $\pazocal{O}(n^3)$, use the Floyd-Warshall algorithm to compute matrix $H^*\in\{0,1\}^{n\times n}$, whose entry $(i,j)$ is $1$ if there is a path $p'$ in $\graph$ from node $j$ to node $i$ and $0$ otherwise.
Apply the Floyd-Warshall algorithm on graph $\graph_{\dint{-n,n}}$ (which consists of $(2n+1)n$ nodes), in time $\pazocal{O}([(2n+1)n]^3) = \pazocal{O}(n^6)$, to check if it contains circuits with positive weight; if yes, then $\graph_\N$ contains an $\infty$-weight path.
If no, the same call of the Floyd-Warshall algorithm provides, for all $z\in\{1,2\}$, $i_z\in\dint{1,n}$, $s_1\in\dint{1,n}$, $s_2\in\dint{-n,-1}$, the largest weight $w(i_z,s_z)=\weight(\hat{q}_z)$ of all pseudo-circuits $\hat{q}_z$ in $\graph_{\dint{-n,n}}$ with $\up(\hat{q}_z)=(i_z,0)$ and $\down(\hat{q}_z)=(i_z,s_z)$ (if such pseudo-circuit $\hat{q}_z$ does not exist, then set $w(i_z,s_z)=-\infty$).
%This computation can be done in time $\pazocal{O}([(2n+1)n]^3) = \pazocal{O}(n^6)$ by applying the Floyd-Warshall algorithm to $G_{\dint{-n,n}}$.
For all tuples $(i_1,i_2,s_1,s_2)$, check if $H^*_{i_2,i_1}=1$, $w(i_1,s_1),w(i_2,s_2)\neq -\infty$, and
\[
    -s_2 w(i_1,s_1)+s_1 w(i_2,s_2)>0 .
\]
This step of the algorithm takes $\pazocal{O}(n^4)$ operations.
The graph $\graph_\N$ contains an $\infty$-weight path if and only if, for at least one tuple, the three conditions are satisfied.
}

We postpone the illustration of the algorithm on an example to the next section, after the introduction of some useful algebraic tools.
\section{An algebra for (infinite) graphs}\label{se:algebra}

The max-plus algebra is a mathematical framework that allows to translate graph-theoretical algorithms for the longest-path problem, such as the Bellman-Ford and Floyd-Warshall algorithms, into algebraic expressions.
Typically, the considered graphs are finite, but in this section we show that extending some results to infinite graphs is possible.

\subsection{Basic algebraic tools}

Before introducing the max-plus algebra, we recall the definition of idempotent semiring.
An \textit{idempotent semiring} $(\D,\oplus,\otimes)$ consists of a set $\D$ endowed with operations $\oplus$ (called addition), which is commutative, associative, idempotent (i.e., $a\oplus a = a$), and has neutral element $\varepsilon$, and $\otimes$ (called multiplication), which is associative, distributive over $\oplus$, has neutral element $e$, and $\forall a\in\D$, $a\otimes \varepsilon=\varepsilon\otimes a = \varepsilon$.
Any idempotent semiring is closed under finite additions and multiplications; if it is also closed under infinite additions and if $\otimes$ distributes over infinite additions, then we say that it is \textit{complete}.
In this case, given any $a\in\D$, the operator $ ^+$ applied to $a$ is defined by $a^+ = \bigoplus_{i\in\N} a^i$, where $a^1=a$ and $a^{i+1} = a\otimes a^i$.
The \textit{Kleene star} of $a$, $a^* = a^+ \oplus e$, satisfies the following property: $a^*\otimes a^* = a^*$.

The max-plus algebra $(\Qbar,\oplus,\otimes)$, where the operations $\oplus$ and $\otimes$ are defined for all $a,b\in\Qbar$ by $a\oplus b=\max\{a,b\}$ and
\[
    a\otimes b = \begin{dcases}
        a+b & \mbox{if } a\neq -\infty \mbox{ and } b\neq -\infty,\\
        -\infty & \mbox{otherwise,}
    \end{dcases}
\]
is a complete idempotent semiring.
\rewrite{On the other hand, $(\Qmax,\oplus,\otimes)$ is an example of idempotent semiring that is not complete.}{}{}
\rerewrite{}{On the other hand, $(\Qmax,\oplus,\otimes)$ is an example of idempotent semiring that is not complete.}
The operations of the max-plus algebra can be extended to matrices of finite and infinite dimensions.
Let $I,J\subseteq \N$; a matrix $A$ is a function $A:I\times J\rightarrow \Qbar$, where $A(i,j)$, denoted by $A_{ij}$, is an entry of $A$.
The collection of such matrices is denoted by $\Qbar^{I\times J}$ or, when $I=\dint{1,m}$ and $J=\dint{1,n}$, with $m,n\in\N$, by $\Qbar^{m\times n}$.
The set $\Qbar^{I\times \{1\}}$ of column vectors is simply indicated by $\Qbar^I$ (or $\Qbar^n$ when $I=\dint{1,n}$).
Given $A,B\in\Qbar^{I_1\times I_2}$, $C\in\Qbar^{I_2\times I_3}$, for all $i\in I_1$, $j\in I_2$, $h\in I_3$, we set
\[
    (A\oplus B)_{ij} = A_{ij} \oplus B_{ij},\quad (A\otimes C)_{ih} = \bigoplus_{k\in I_2} A_{ik}\otimes C_{kh}.
\]
With these definitions, $(\Qbar^{n\times n},\oplus,\otimes)$ and $(\Qbar^{\N\times \N},\oplus,\otimes)$ form two complete idempotent semirings, see \cite[Section 1.4]{droste2009handbook}.
In such semirings, the neutral element for $\oplus$ is the matrix $\pazocal{E}$ whose entries are all $-\infty$, and the neutral element for $\otimes$ is the matrix $E$ such that $E_{ii}=0$ for all $i$ and $E_{ij} = -\infty$ for all $i\neq j$.
Given two matrices $A$ and $B$ of the same size, we write $A\leq B$ iff, for all $i,j$, $A_{ij}\leq B_{ij}$.
Note that this is equivalent to $A\oplus B = B$.

\begin{thm}{\cite[Theorem 4.70]{baccelli1992synchronization}}\label{th:Kleene_expressions}
Let $A\in\Qbar^{I\times I}$ and $x\in\Qbar^{I}$.
Then $x \geq A\otimes x$ iff $x = A^* \otimes x$.\myendtheoremsymbol
\end{thm}

\subsection{Max-plus algebra and $\N$-periodic graphs}

Let $\graph=(\nodes,\arcs,w)$ be a (possibly infinite) weighted directed graph with node set $\nodes$, arc set $\arcs$, and weight function $w$.
We say that matrix $A\in\Qmax^{I\times I}$ is the \textit{adjacency matrix} of $\graph$, and that $\graph$ is the \textit{precedence graph} of $A$, if $\nodes=I$ and there is an arc in $\graph$ from node $i$ to node $j$ of weight $w((i,j))=A_{ji}$ if and only if $A_{ji}\neq -\infty$.
In this case, we write $\graph = \graph(A)$.
Furthermore, $(A^m)_{ij}$ and $(A^+)_{ij}$ correspond, respectively, to the largest weight of all paths in $\graph(A)$ from node $j$ to node $i$ of length $m$ and of any length.
Therefore, $A^*=A^+\oplus E\in\Qmax^{I\times I}$ if and only if there are no $\infty$-weight paths in $\graph(A)$; otherwise, at least one entry in $A^*$ is equal to $+\infty$.

Take an $\N$-periodic graph $\graph_\N=\graph_\N(M_{-1},M_0,M_{+1})$, where $M_{s}\in\Qmax^{n\times n}$ for all $s\in\{-1,0,+1\}$, and rename each node $(i,k)$ of $\graph_\N$ into $i+(k-1)n\in\N$.
In this way, we get an isomorphic graph, say $\graph(M)$, with node set $\N$.
From the definition of $\N$-periodic graphs, its adjacency matrix $M$ has the following block-tridiagonal form:
\begin{equation}\label{eq:block_tridiagonal}
    M = \begin{bsmallmatrix}
        M_0 & M_{-1} & \pazocal{E} & \pazocal{E} & \pazocal{E} &  \cdots\\
        M_{+1} & M_{0} & M_{-1} & \pazocal{E} & \pazocal{E} &  \cdots\\
        \pazocal{E} & M_{+1} & M_{0} & M_{-1} & \pazocal{E} &  \cdots\\
        \pazocal{E} & \pazocal{E} & M_{+1} & M_{0} & M_{-1} &  \cdots\\
        \pazocal{E} & \pazocal{E} & \pazocal{E} & M_{+1} & M_{0}  & \cdots\\
        %\pazocal{E} & \pazocal{E} & \pazocal{E} & \pazocal{E} & M_{+1} & M_{0} & \cdots\\
        \svdots & \svdots & \svdots & \svdots & \svdots & \sddots
    \end{bsmallmatrix}\in\Qmax^{\N\times \N}.
\end{equation}
Therefore, an equivalent algebraic formulation of Theorem~\ref{th:polynomial_complexity} is that, given a block-tridiagonal matrix $M\in\Qmax^{\N\times \N}$ of the form~\eqref{eq:block_tridiagonal}, it is possible to decide in time $\pazocal{O}(n^6)$ whether $M^*\in\Qmax^{\N\times \N}$.
This is illustrated in the following example.

\begin{exmp}\label{ex:periodic_graphs_max_plus}
Consider the $\N$-periodic graph $\graph_\N$ of Figure~\ref{fi:periodic_graph}.
The corresponding adjacency matrix $M$ has the form~\eqref{eq:block_tridiagonal}, where $M_{-1},M_0,M_{+1}$ have been defined in Example~\ref{ex:simple_graph}.
From the informal discussion carried out in Example~\ref{ex:simple_graph_2}, we know that $\graph_\N$ contains an $\infty$-weight path only for certain parameters $\alpha,\beta$.
We can now verify this using the algorithm outlined in Section~\ref{se:N_periodic_graphs}.

To start, we compute the matrix $H^*=\begin{bsmallmatrix}
    1 & 0\\1 & 1
\end{bsmallmatrix}$ using the Floyd-Warshall algorithm.
The only $0$-entry in position $(1,2)$ of $H^*$ represents the fact that there are no paths from node $2$ to node $1$ in the static graph $\graph$ generating $\graph_\N$ (see Figure~\ref{fi:periodic_graph}).
Let $\graph(A)$ be the precedence graph isomorphic to $\graph_{\dint{-2,2}}$ obtained by renaming each node $(i,k)\in\{1,2\}\times\dint{-2,2}$ into $i+2(k+2)\in\dint{1,10}$.
Then the corresponding adjacency matrix is
\[
	%\begin{array}{rl}
	A \!= \!\!%&
	\begin{bsmallmatrix}
		M_0 & M_{-1} & \pazocal{E} & \pazocal{E} & \pazocal{E} \\ 
		M_{+1} & M_0 & M_{-1} & \pazocal{E} & \pazocal{E} \\ 
		\pazocal{E} & M_{+1} & M_0 & M_{-1} & \pazocal{E} \\ 
		\pazocal{E} & \pazocal{E} & M_{+1} & M_0 & M_{-1} \\ 
		\pazocal{E} & \pazocal{E} & \pazocal{E} & M_{+1} & M_0 \\ 
	\end{bsmallmatrix}%\\
	\!\!=\!\!%&
	\begin{bsmallmatrix}
	\cdot&\cdot&\alpha  &\cdot  &\cdot  &\cdot  &\cdot  &\cdot  &\cdot  &\cdot\\
	0  &\cdot  &\cdot    &-3  &\cdot  &\cdot  &\cdot  &\cdot  &\cdot  &\cdot\\
	\beta  &\cdot  &\cdot  &\cdot    &\alpha  &\cdot  &\cdot  &\cdot  &\cdot  &\cdot\\
	\cdot     &2     &0  &\cdot  &\cdot    &-3  &\cdot  &\cdot  &\cdot  &\cdot\\
	\cdot  &\cdot     &\beta  &\cdot  &\cdot  &\cdot    &\alpha  &\cdot  &\cdot  &\cdot\\
	\cdot  &\cdot  &\cdot     &2     &0  &\cdot  &\cdot    &-3  &\cdot  &\cdot\\
	\cdot  &\cdot  &\cdot  &\cdot     &\beta  &\cdot  &\cdot  &\cdot    &\alpha  &\cdot\\
	\cdot  &\cdot  &\cdot  &\cdot  &\cdot     &2     &0  &\cdot  &\cdot    &-3\\
	\cdot  &\cdot  &\cdot  &\cdot  &\cdot  &\cdot     &\beta  &\cdot  &\cdot  &\cdot\\
	\cdot  &\cdot  &\cdot  &\cdot  &\cdot  &\cdot  &\cdot     &2     &0  &\cdot
	\end{bsmallmatrix}%\in\Qmax^{10\times 10},
	%\begin{bsmallmatrix}
	%-\infty&-\infty&-5  &-\infty  &-\infty  &-\infty  &-\infty  &-\infty  &-\infty  &-\infty\\
	%0  &-\infty  &-\infty    &-3  &-\infty  &-\infty  &-\infty  &-\infty  &-\infty  &-\infty\\
	%4  &-\infty  &-\infty  &-\infty    &-5  &-\infty  &-\infty  &-\infty  &-\infty  &-\infty\\
	%-\infty     &2     &0  &-\infty  &-\infty    &-3  &-\infty  &-\infty  &-\infty  &-\infty\\
	%-\infty  &-\infty     &4  &-\infty  &-\infty  &-\infty    &-5  &-\infty  &-\infty  &-\infty\\
	%-\infty  &-\infty  &-\infty     &2     &0  &-\infty  &-\infty    &-3  &-\infty  &-\infty\\
	%-\infty  &-\infty  &-\infty  &-\infty     &4  &-\infty  &-\infty  &-\infty    &-5  &-\infty\\
	%-\infty  &-\infty  &-\infty  &-\infty  &-\infty     &2     &0  &-\infty  &-\infty    &-3\\
	%-\infty  &-\infty  &-\infty  &-\infty  &-\infty  &-\infty     &4  &-\infty  &-\infty  &-\infty\\
	%-\infty  &-\infty  &-\infty  &-\infty  &-\infty  &-\infty  &-\infty     &2     &0  &-\infty
	%\end{bsmallmatrix}
	%\end{array}
\]
where each "$\cdot$" represents a $-\infty$.
Applying the Floyd-Warshall algorithm to graph $\graph_{\dint{-2,2}}$ is equivalent to computing the Kleene star of $A$.

For parameters $\alpha=-1,\beta=2$, $A^*$ contains entries equal to $+\infty$; thus, there exists a circuit in $\graph_\N$ with positive weight, and $M^*\notin\Qmax^{\N\times \N}$.

When $\alpha=-5,\beta=4$, we get
\setcounter{MaxMatrixCols}{11}
\[
	A^* \!\! = \!\!
	\begin{bNiceArray}{cccccccccc}[small,first-row,last-col,code-for-first-row=\scriptscriptstyle,code-for-last-col=\scriptscriptstyle]
	(1,-2)&(2,-2)&(1,-1)&(2,-1)&(1,0)&(2,0)&(1,1)&(2,1)&(1,2)&(2,2)&(i,k)\\
	0  &\cdot    &-5  &\cdot   &-10  &\cdot   &-15  &\cdot   &-20  &\cdot&(1,-2)\\
	4     &0     &0    &-3    &-4    &-6    &-8    &-9   &-12   &-12&(2,-2)\\
	4  &\cdot     &0  &\cdot    &-5  &\cdot   &-10  &\cdot   &-15  &\cdot&(1,-1)\\
	7     &2     &3     &0    &-1    &-3    &-5    &-6    &-9    &-9&(2,-1)\\
	8  &\cdot     &4  &\cdot     &0  &\cdot    &-5  &\cdot   &-10  &\cdot&(1,0)\\
    10     &4     &6     &2     &2     &0    &-2    &-3    &-6    &-6&(2,0)\\
    12  &\cdot     &8  &\cdot     &4  &\cdot     &0  &\cdot    &-5  &\cdot&(1,1)\\
    13     &6     &9     &4     &5     &2     &1     &0    &-3    &-3&(2,1)\\
    16  &\cdot    &12  &\cdot     &8  &\cdot     &4  &\cdot     &0  &\cdot&(1,2)\\
    16     &8    &12     &6     &8     &4     &4     &2     &0     &0&(2,2)
	\CodeAfter
	%\tikz\draw[gray] ($(1-|5)+(0,.02)$)--($(1-|7)+(0,.02)$)--(11-|7)--(11-|5)--cycle;
	\tikz\draw[gray] ($(1-|5)+(0.0,.03)$)--($(1-|6)+(0.0,.03)$)--($(2-|6)+(.0,.03)$)--($(2-|5)+(.00,.03)$)--cycle;
	\tikz\draw[gray] ($(3-|5)+(0.0,.03)$)--($(3-|6)+(0.0,.03)$)--($(4-|6)+(.0,.03)$)--($(4-|5)+(.00,.03)$)--cycle;
	\tikz\draw[myblue,densely dotted,thick] ($(7-|5)+(0.0,.03)$)--($(7-|6)+(0.0,.03)$)--($(8-|6)+(.0,.03)$)--($(8-|5)+(.00,.03)$)--cycle;
	\tikz\draw[gray] ($(9-|5)+(0.0,.03)$)--($(9-|6)+(0.0,.03)$)--($(10-|6)+(.0,.03)$)--($(10-|5)+(.00,.03)$)--cycle;
	\tikz\draw[gray] ($(2-|6)+(0.03,.03)$)--($(2-|7)+(0.00,.03)$)--($(3-|7)+(.0,.03)$)--($(3-|6)+(.03,.03)$)--cycle;
	\tikz\draw[myred,densely dashed,thick] ($(4-|6)+(0.03,.03)$)--($(4-|7)+(0.00,.03)$)--($(5-|7)+(.0,.03)$)--($(5-|6)+(.03,.03)$)--cycle;
	\tikz\draw[gray] ($(8-|6)+(0.03,.03)$)--($(8-|7)+(0.00,.03)$)--($(9-|7)+(.0,.03)$)--($(9-|6)+(.03,.03)$)--cycle;
	\tikz\draw[gray] ($(10-|6)+(0.03,.03)$)--($(10-|7)+(0.00,.03)$)--($(11-|7)+(.0,.03)$)--($(11-|6)+(.03,.03)$)--cycle;
	\end{bNiceArray}.
\]
\rerewrite{Since no $+\infty$ is present in $A^*$, $\graph_{\dint{-2,2}}$ and, consequently, $\graph_\N$ do not contain circuits with positive weight and length $\leq n=2$.}{Since no $+\infty$ is present in $A^*$, $\graph_{\dint{-2,2}}$ does not contain circuits with positive weight.}
\rerewrite{
Each of the eight entries framed in the above representation of $A^*$ corresponds to a circuit with length at most $n$ of the static graph $\graph$.
From the position and value of such entries, we can extract the shift, weight, and source node of the associated circuits.
For example, we infer that there is a circuit $q_1$ with shift $s_1=1>0$, weight $w_1=4$, and source node $i_1=1$ (see the element of $A^*$ in the \textcolor{myblue}{\bf blue}, dotted frame), and a circuit $q_2$ with shift $s_2=-1<0$, weight $w_2=-3$, and source node $i_2=2$ (see the element of $A^*$ in the \textcolor{myred}{\bf red}, dashed frame).
}{
Each of the eight entries framed in the above representation of $A^*$ corresponds to a pseudo-circuit of $\graph_{\dint{-2,2}}$.
From the position and value of such entries, we can extract their shift, weight, and source node.
For example, we infer that there is a pseudo-circuit $q_1$ with shift $s_1=1>0$, weight $w_1=4$, and source node $i_1=1$ (see the element of $A^*$ in the \textcolor{myblue}{\bf blue}, dotted frame), and a pseudo-circuit $q_2$ with shift $s_2=-1<0$, weight $w_2=-3$, and source node $i_2=2$ (see the element of $A^*$ in the \textcolor{myred}{\bf red}, dashed frame).
}
Since $(H^*)_{21} = 1$, there exists a path $p'$ in $\graph$ from node $1$ to node $2$.
Moreover, $-s_2 w_1 + s_1 w_2 = 1>0$.
\rerewrite{
We can then conclude that a path $p=q_1 p' q_2$ that satisfies the conditions of Theorem~\ref{le:main} exists.
Thus, $\graph_\N$ contains an $\infty$-weight path and $M^*$ has entries equal to $+\infty$.
}{
We can then conclude that $\graph_\N$ contains an $\infty$-weight path and $M^*$ has entries equal to $+\infty$.
}

Finally, considering values $\alpha=-1,\beta=1$, we obtain
\[
	A^* \!\! = \!\!
	\begin{bNiceArray}{cccccccccc}[small,first-row,last-col,code-for-first-row=\scriptscriptstyle,code-for-last-col=\scriptscriptstyle]
	(1,-2)&(2,-2)&(1,-1)&(2,-1)&(1,0)&(2,0)&(1,1)&(2,1)&(1,2)&(2,2)&(i,k)\\
    0&  \cdot&    -1&  \cdot&    -2&  \cdot&    -3&  \cdot&    -4&  \cdot&(1,-2)\\
    0&     0&    -1&    -3&    -2&    -6&    -3&    -9&    -4&   -12&(2,-2)\\
    1&  \cdot&     0&  \cdot&    -1&  \cdot&    -2&  \cdot&    -3&  \cdot&(1,-1)\\
    2&     2&     1&     0&     0&    -3&    -1&    -6&    -2&    -9&(2,-1)\\
    2&  \cdot&     1&  \cdot&     0&  \cdot&    -1&  \cdot&    -2&  \cdot&(1,0)\\
    4&     4&     3&     2&     2&     0&     1&    -3&     0&    -6&(2,0)\\
    3&  \cdot&     2&  \cdot&     1&  \cdot&     0&  \cdot&    -1&  \cdot&(1,1)\\
    6&     6&     5&     4&     4&     2&     3&     0&     2&    -3&(2,1)\\
    4&  \cdot&     3&  \cdot&     2&  \cdot&     1&  \cdot&     0&  \cdot&(1,2)\\
    8&     8&     7&     6&     6&     4&     5&     2&     4&    0 &(2,2)
	\CodeAfter
	\tikz\draw[gray] ($(1-|5)+(0.0,.03)$)--($(1-|6)+(0.0,.03)$)--($(2-|6)+(.0,.03)$)--($(2-|5)+(.00,.03)$)--cycle;
	\tikz\draw[gray] ($(3-|5)+(0.0,.03)$)--($(3-|6)+(0.0,.03)$)--($(4-|6)+(.0,.03)$)--($(4-|5)+(.00,.03)$)--cycle;
	\tikz\draw[gray] ($(7-|5)+(0.0,.03)$)--($(7-|6)+(0.0,.03)$)--($(8-|6)+(.0,.03)$)--($(8-|5)+(.00,.03)$)--cycle;
	\tikz\draw[gray] ($(9-|5)+(0.0,.03)$)--($(9-|6)+(0.0,.03)$)--($(10-|6)+(.0,.03)$)--($(10-|5)+(.00,.03)$)--cycle;
	\tikz\draw[gray] ($(2-|6)+(0.03,.03)$)--($(2-|7)+(0.00,.03)$)--($(3-|7)+(.0,.03)$)--($(3-|6)+(.03,.03)$)--cycle;
	\tikz\draw[gray] ($(4-|6)+(0.03,.03)$)--($(4-|7)+(0.00,.03)$)--($(5-|7)+(.0,.03)$)--($(5-|6)+(.03,.03)$)--cycle;
	\tikz\draw[gray] ($(8-|6)+(0.03,.03)$)--($(8-|7)+(0.00,.03)$)--($(9-|7)+(.0,.03)$)--($(9-|6)+(.03,.03)$)--cycle;
	\tikz\draw[gray] ($(10-|6)+(0.03,.03)$)--($(10-|7)+(0.00,.03)$)--($(11-|7)+(.0,.03)$)--($(11-|6)+(.03,.03)$)--cycle;
	\end{bNiceArray}.
\]
\rerewrite{
As in the previous case, we can conclude that there are no circuits with positive weight and length at most $n=2$ in $\graph_\N$.
Moreover, analyzing each of the eight circuits corresponding to the framed entries of $A^*$, it is not possible to find circuits $q_1,q_2$ with shifts $s_1>0,s_2<0$ and weights $w_1,w_2$ such that $-s_2w_1+s_1w_2>0$.
Therefore, $\graph_\N$ does not contain $\infty$-weight paths and $M^*\in\Qmax^{\N\times \N}$.
}{
As in the previous case, there are no circuits with positive weight in $\graph_{\dint{-2,2}}$.
Moreover, analyzing each of the eight pseudo-circuits corresponding to the framed entries of $A^*$, it is not possible to find pseudo-circuits $q_1,q_2$ with shifts $s_1>0,s_2<0$ and weights $w_1,w_2$ such that $-s_2w_1+s_1w_2>0$.
Therefore, $\graph_\N$ does not contain $\infty$-weight paths and $M^*\in\Qmax^{\N\times \N}$.
}
\myenddefinitionsymbol
\end{exmp}

There is an important connection between precedence graphs and linear inequalities in the max-plus algebra, which is given by the following proposition.
\begin{prop}[\cite{butkovivc2010max}]
Given a matrix $A\in\Qmax^{n\times n}$, the inequality $x\geq A\otimes x$ admits a rational solution $x\in\Q^n$ iff $\graph(A)$ does not contain positive-weight circuits.\myendtheoremsymbol
\end{prop}

\rewrite{We shall now prove that the latter proposition extends to the case of $\N$-periodic graphs.}{The following result, proven in \zor{CITE ARXIV PAPER}, shows that the latter proposition extends to the case of $\N$-periodic graphs.}{}

\rerewrite{}{We shall now prove that the latter proposition extends to the case of $\N$-periodic graphs.}

\begin{prop}\label{pr:infinite_matrix_inequalities}
Given a matrix $M\in\Qmax^{\N\times \N}$ of the form~\eqref{eq:block_tridiagonal}, the inequality $x\geq M\otimes x$ admits a rational solution $x\in\Q^\N$ iff $\graph(M)$ does not contain $\infty$-weight paths. \myendtheoremsymbol
\end{prop}
\rewrite{
\begin{pf}
Recall that $\graph(M)$ contains $\infty$-weight paths iff $M^*$ has entries equal to $+\infty$.

"$\Rightarrow$": this part of the proof is by contrapositive.
Suppose that $(M^*)_{ij}=+\infty$ for some $i,j\in\N$.
Then, from Theorem~\ref{th:Kleene_expressions}, the $i$-th inequality of $x\geq M\otimes x$ is equivalent to equation
\[
    x_i = \left(\bigoplus_{h\neq j} (M^*)_{ih}\otimes x_h\right) \oplus +\infty\otimes x_j.
\]
If $x_j\in\Q$, the latter expression implies that $x_i = +\infty$.
Thus, there is no rational solution of $x\geq M\otimes x$.

"$\Leftarrow$": this part of the proof is constructive.
Suppose that $M^*\in\Qmax^{\N\times \N}$.
In the following, we provide a specific solution of $x\geq M\otimes x$ and then show that it is rational.
We will use the notation
\[
    M^* = \begin{bsmallmatrix}
        \pazocal{M}_{1,1} & \pazocal{M}_{1,2} & \pazocal{M}_{1,3} & \pazocal{M}_{1,4} & \cdots\\
        \pazocal{M}_{2,1} & \pazocal{M}_{2,2} & \pazocal{M}_{2,3} & \pazocal{M}_{2,4} & \cdots\\
        \pazocal{M}_{3,1} & \pazocal{M}_{3,2} & \pazocal{M}_{3,3} & \pazocal{M}_{3,4} & \cdots\\
        \pazocal{M}_{4,1} & \pazocal{M}_{4,2} & \pazocal{M}_{4,3} & \pazocal{M}_{4,4} & \cdots\\
        \svdots & \svdots & \svdots & \svdots & \sddots
    \end{bsmallmatrix} = \begin{bsmallmatrix}
        \pazocal{M}_1 & \pazocal{M}_2 & \pazocal{M}_3 & \pazocal{M}_4 & \cdots
    \end{bsmallmatrix},
\]
where $\pazocal{M}_{i,j}\in\Qmax^{n\times n}$ and $\pazocal{M}_{j}\in\Qmax^{\N,\dint{1,n}}$.
Define
\[
    x = \bigoplus_{k\in\N} \pazocal{M}_k \otimes \alpha_k,
\]
where $\alpha_k$ is an $n$-dimensional vector whose entries are all equal to
\[
    -\max\{(\pazocal{M}_{1,h})_{ij}\mid i,j\in\dint{1,n},\ h\in\dint{1,k}\}.
    %\alpha_k = - \max_{i,j\in\dint{1,n}} \max_{h\in\dint{1,k}} (\pazocal{M}_{1,h})_{ij}.
    %\left(\bigoplus_{h=1}^{k} \| \pazocal{M}_{1,k} \| \right)^{-1}.
\]

From the distributivity of $\otimes$ over infinite $\oplus$ in $(\Qbar,\oplus,\otimes)$ and the fact that $M^*\otimes M^* = M^*$, we have
\[
    \begin{array}{rcl}
    M^* \otimes x &=& \displaystyle M^* \otimes \left(\bigoplus_{k\in\N} \pazocal{M}_k \otimes \alpha_k\right)\\
    &=& \displaystyle \bigoplus_{k\in\N} M^* \otimes \pazocal{M}_k \otimes \alpha_k
    = \bigoplus_{k\in\N} \pazocal{M}_k \otimes \alpha_k =x.
    \end{array}
\]
Due to Theorem~\ref{th:Kleene_expressions}, this implies that $x$ is a solution of $x\geq M\otimes x$.

To conclude the proof, we need to show that $x\in\Q^\N$.
In the following, we write $x = [x^{1\top}\ x^{2\top}\ \cdots]^\top$, where $x^h = \bigoplus_{k\in\N} \pazocal{M}_{h,k}\otimes \alpha_k\in\Qbar^n$.
First of all, observe that $\alpha_k\in\Q^n$, as $\pazocal{M}_{1,1}\geq E$. 
Hence, $x\in(\Q\cup\{+\infty\})^\N$, since
\[
    x^h
    = 
    \bigoplus_{k\in\N}
        \pazocal{M}_{h,k} \otimes \alpha_k
    \geq 
        \pazocal{M}_{h,h} \otimes \alpha_h
    \geq
        E \otimes \alpha_h=
        \alpha_h.
\]
To show that $x^h$ does not contain $+\infty$'s, we rewrite it as
\[
    x^h = \bigoplus_{k\leq h} \pazocal{M}_{h,k} \otimes \alpha_k \oplus \bigoplus_{k > h} \pazocal{M}_{h,k} \otimes \alpha_k.
\]
From the closure of $\Qmax$ under finitely many $\oplus$ and $\otimes$, $\bigoplus_{k\leq h} \pazocal{M}_{h,k} \otimes \alpha_k\in\Qmax^{n}$; therefore, we only need to show that $\bigoplus_{k > h} \pazocal{M}_{h,k} \otimes \alpha_k\in\Qmax^{n}$.
Let $h<k\in\N$, $i,j\in\dint{1,n}$, and let $p$ be a path of maximum weight in the $\N$-periodic graph isomorphic to $\graph(M)$ from node $(i,k)$ to node $(j,h)$, i.e., $\weight(p) = (\pazocal{M}_{h,k})_{ji}$.
The path $p$ can be factored into $p = p'q$, where $p'$ is a maximum-weight path from $(i,k)$ to some node $(j',h)$ that never visits a node with shift $h-1$, and $q$ is a (possibly empty) maximum-weight path from $(j',h)$ to $(j,h)$.
The periodicity of $\N$-periodic graphs implies that $\weight(p') = (\pazocal{M}_{1,k-h})_{j'i}$.
Similarly, $\weight(q) = (\pazocal{M}_{h,h})_{jj'}$.
This reasoning shows that, for all $h<k$, $\pazocal{M}_{h,k} = \pazocal{M}_{h,h} \otimes \pazocal{M}_{1,k-h}$.
Since by definition each element of $\alpha_k$ is less than or equal to $-(\pazocal{M}_{1,k-h})_{ij}$ for all $i,j$, we have
\[
    \bigoplus_{k>h} \pazocal{M}_{h,k}\otimes \alpha_k = \bigoplus_{k>h}\pazocal{M}_{h,h}\otimes \pazocal{M}_{1,k-h} \otimes \alpha_k \leq \pazocal{M}_{h,h}\otimes \vec{e},
\]
where $\vec{e}$ is the $n$-dimensional vector with only $0$'s.
As $\pazocal{M}_{h,h}\otimes \vec{e}\in\Qmax^{n}$, we conclude that $\bigoplus_{k>h}\pazocal{M}_{h,k}\otimes \alpha_k$, and thus $x^h$, are both elements of $\Qmax^n$. \myendproofsymbol
\end{pf}
}{}{}
\rerewrite{}{
\begin{pf}
Recall that $\graph(M)$ contains $\infty$-weight paths iff $M^*$ has entries equal to $+\infty$.

"$\Rightarrow$": this part of the proof is by contrapositive.
Suppose that $(M^*)_{ij}=+\infty$ for some $i,j\in\N$.
Then, from Theorem~\ref{th:Kleene_expressions}, the $i$-th inequality of $x\geq M\otimes x$ is equivalent to equation
    $x_i = \left(\bigoplus_{h\neq j} (M^*)_{ih}\otimes x_h\right) \oplus +\infty\otimes x_j$.
If $x_j\in\Q$, the latter expression implies that $x_i = +\infty$.
Thus, there is no rational solution of $x\geq M\otimes x$.

"$\Leftarrow$": this part of the proof is constructive.
Suppose that $M^*\in\Qmax^{\N\times \N}$.
In the following, we provide a specific solution of $x\geq M\otimes x$ and then show that it is rational.
We will use the notation
\[
    M^* = \begin{bsmallmatrix}
        \pazocal{M}_{1,1} & \pazocal{M}_{1,2} & \pazocal{M}_{1,3} & \pazocal{M}_{1,4} & \cdots\\
        \pazocal{M}_{2,1} & \pazocal{M}_{2,2} & \pazocal{M}_{2,3} & \pazocal{M}_{2,4} & \cdots\\
        \pazocal{M}_{3,1} & \pazocal{M}_{3,2} & \pazocal{M}_{3,3} & \pazocal{M}_{3,4} & \cdots\\
        \pazocal{M}_{4,1} & \pazocal{M}_{4,2} & \pazocal{M}_{4,3} & \pazocal{M}_{4,4} & \cdots\\
        \svdots & \svdots & \svdots & \svdots & \sddots
    \end{bsmallmatrix} = \begin{bsmallmatrix}
        \pazocal{M}_1 & \pazocal{M}_2 & \pazocal{M}_3 & \pazocal{M}_4 & \cdots
    \end{bsmallmatrix},
\]
where $\pazocal{M}_{i,j}\in\Qmax^{n\times n}$ and $\pazocal{M}_{j}\in\Qmax^{\N,\dint{1,n}}$.
Define
\[
    x = \bigoplus_{k\in\N} \pazocal{M}_k \otimes \alpha_k,
\]
where $\alpha_k$ is an $n$-dimensional vector whose entries are all equal to
\[
    -\max\{(\pazocal{M}_{1,h})_{ij}\mid i,j\in\dint{1,n},\ h\in\dint{1,k}\}.
    %\alpha_k = - \max_{i,j\in\dint{1,n}} \max_{h\in\dint{1,k}} (\pazocal{M}_{1,h})_{ij}.
    %\left(\bigoplus_{h=1}^{k} \| \pazocal{M}_{1,k} \| \right)^{-1}.
\]

From the distributivity of $\otimes$ over infinite $\oplus$ in $(\Qbar,\oplus,\otimes)$ and the fact that $M^*\otimes M^* = M^*$, we have
\[
    \begin{array}{rcl}
    M^* \otimes x &=& \displaystyle M^* \otimes \left(\bigoplus_{k\in\N} \pazocal{M}_k \otimes \alpha_k\right)\\
    &=& \displaystyle \bigoplus_{k\in\N} M^* \otimes \pazocal{M}_k \otimes \alpha_k
    = \bigoplus_{k\in\N} \pazocal{M}_k \otimes \alpha_k =x.
    \end{array}
\]
Due to Theorem~\ref{th:Kleene_expressions}, this implies that $x$ is a solution of $x\geq M\otimes x$.

To conclude the proof, we need to show that $x\in\Q^\N$.
In the following, we write $x = [x^{1\top}\ x^{2\top}\ \cdots]^\top$, where $x^h = \bigoplus_{k\in\N} \pazocal{M}_{h,k}\otimes \alpha_k\in\Qbar^n$.
First of all, observe that $\alpha_k\in\Q^n$, as $\pazocal{M}_{1,1}\geq E$. 
Hence, $x\in(\Q\cup\{+\infty\})^\N$, since
\[
    x^h
    = 
    \bigoplus_{k\in\N}
        \pazocal{M}_{h,k} \otimes \alpha_k
    \geq 
        \pazocal{M}_{h,h} \otimes \alpha_h
    \geq
        E \otimes \alpha_h=
        \alpha_h.
\]
To show that $x^h$ does not contain $+\infty$'s, we rewrite it as
\[
    x^h = \bigoplus_{k\leq h} \pazocal{M}_{h,k} \otimes \alpha_k \oplus \bigoplus_{k > h} \pazocal{M}_{h,k} \otimes \alpha_k.
\]
From the closure of $\Qmax$ under finitely many $\oplus$ and $\otimes$, $\bigoplus_{k\leq h} \pazocal{M}_{h,k} \otimes \alpha_k\in\Qmax^{n}$; therefore, we only need to show that $\bigoplus_{k > h} \pazocal{M}_{h,k} \otimes \alpha_k\in\Qmax^{n}$.
Let $h<k\in\N$, $i,j\in\dint{1,n}$, and let $p$ be a path of maximum weight in the $\N$-periodic graph isomorphic to $\graph(M)$ from node $(i,k)$ to node $(j,h)$, i.e., $\weight(p) = (\pazocal{M}_{h,k})_{ji}$.
Such a path $p$ is always formed by concatenating a maximum-weight path $p'$ from $(i,k)$ to some node $(j',h)$ that never visits a node with shift $h-1$, and a (possibly empty) maximum-weight path $q$ from $(j',h)$ to $(j,h)$.
The periodicity of $\N$-periodic graphs implies that $\weight(p') = (\pazocal{M}_{1,k-h})_{j'i}$.
Similarly, $\weight(q) = (\pazocal{M}_{h,h})_{jj'}$.
This reasoning shows that, for all $h<k$, $\pazocal{M}_{h,k} = \pazocal{M}_{h,h} \otimes \pazocal{M}_{1,k-h}$.
Since by definition each element of $\alpha_k$ is less than or equal to $-(\pazocal{M}_{1,k-h})_{ij}$ for all $i,j$, we have
\[
    \bigoplus_{k>h} \pazocal{M}_{h,k}\otimes \alpha_k = \bigoplus_{k>h}\pazocal{M}_{h,h}\otimes \pazocal{M}_{1,k-h} \otimes \alpha_k \leq \pazocal{M}_{h,h}\otimes \vec{e},
\]
where $\vec{e}$ is the $n$-dimensional vector with only $0$'s.
As $\pazocal{M}_{h,h}\otimes \vec{e}\in\Qmax^{n}$, we conclude that $\bigoplus_{k>h}\pazocal{M}_{h,k}\otimes \alpha_k$, and thus $x^h$, are both elements of $\Qmax^n$. \myendproofsymbol
\end{pf}}
\section{Consistency of P-time event graphs}\label{se:P_TEGs}

In the present section we combine Theorem~\ref{th:polynomial_complexity} and Proposition~\ref{pr:infinite_matrix_inequalities} to analyze P-time event graphs.
We start by recalling their definition and dynamics.

\subsection{P-time event graphs}

\begin{defn}[\cite{khansa1996p}]\label{de:PTPN}
An ordinary \textit{P-time Petri net} is a 5-tuple $(\places,\transitions,\arcs,m,\iota)$, in which $\places$ a finite set of places, $\transitions$ is a finite set of transitions, $\arcs\subseteq (\places\times \transitions)\cup (\transitions \times \places)$ is the set of arcs connecting places to transitions and transitions to places, and $m:\places\rightarrow\No$ and $\iota:\places\rightarrow\{[\tau^-,\tau^+]\cap \Q \mid \tau^-\in \Q_{\geq 0},\tau^+\in\Q_{\geq 0}\cup\{\infty\}\}$ are two maps that associate to each place $p\in\places$, respectively, its initial number of tokens (or marking) $m(p)$, and a time interval $\iota(p)=[\tau_p^-,\tau_p^+]\cap \Q$. \myenddefinitionsymbol
\end{defn}

The dynamics of a P-time Petri net evolves as follows.
A transition $t\in\transitions$ is said to be enabled if either it has no upstream places (i.e., $\forall p\in\places$, $(p,t)\notin\arcs$) or each upstream place $p\in \places$ contains at least one token that has resided in $p$ for a time included in interval $[\tau_p^-,\tau_p^+]\cap\Q$.
Note that this time interval is always closed, unless $\tau_p^+=+\infty$, in which case it is of the form $[\tau_p^-,+\infty)$.
When transition $t$ is enabled, it can fire, causing one token to be instantaneously removed from each upstream place and one token to be instantaneously added to each downstream place.
If a token resides for too long in $p$, violating the constraint imposed by interval $[\tau_p^-,\tau_p^+]\cap \Q$, then the token is said to be \textit{dead}.
%As we will see later, in some P-time Petri nets this event cannot be avoided by carefully choosing the firing time of each transition; consequently, these P-time Petri nets will be called inconsistent.

In this paper, we focus on a subclass of P-time Petri nets called \textit{P-time event graphs} (P-TEGs).
A P-TEG is a P-time Petri net where each place has exactly one upstream and one downstream transition (i.e., $\forall p\in \places$, $\exists ! (t_{\textup{up}},t_{\textup{down}})\in\transitions\times \transitions$ such that $(t_{\textup{up}},p)\in\arcs$ and $(p,t_{\textup{down}})\in\arcs$).
We say that a P-TEG is \emph{consistent} if there exists an infinite sequence of firings of its transitions that does not cause any token death.
In order to study this property, it is convenient to state the dynamics of P-TEGs as a system of inequalities.

\subsection{Dynamics}

\begin{rem}\label{re:transformation}
To simplify the following discussions, we recall from~\cite{amari2005control} that there exists a transformation that, given a P-TEG $(\places,\transitions,\arcs,m,\iota)$, returns another one preserving the behavior of existing transitions, in which the initial number of tokens in every place is either 0 or 1.
If $|\transitions|$ is the number of transitions in the original P-TEG, then the number of transitions in the transformed P-TEG is
%\begin{equation}\label{eq:transform_PTEG}
	$n = |\transitions| + \sum_{p\in\places} \max(0,m(p)-1)$.
%\end{equation}
Assuming unary encoding of $m(p)$, the time complexity of the transformation is therefore polynomial.
Thus, without loss of generality, in the remainder of the paper, only P-TEGs in which every place has either 0 or 1 initial token are considered, i.e., $\forall p\in\places$, $m(p)\in\{0,1\}$.
Note that, in such P-TEGs, $n$ and $|\transitions|$ coincide.  \myenddefinitionsymbol
\end{rem}

The assumption of Remark~\ref{re:transformation} allows to formulate the (nondeterministic) dynamics of a P-TEG $(\places,\transitions,\arcs,m,\iota)$ with $|\transitions| = n$ transitions as the system of linear inequalities presented in the following\footnote{To be precise, in this paper we focus on P-TEGs with loose initial conditions, as defined in~\cite{zorzenon2023switched}. This means that we will not consider the arrival time of the initial tokens as part of the dynamics. The extension of this work to P-TEGs with strict initial conditions is left as future work.}.
Let us define matrices $A^0,A^1\in\Qmax^{n\times n}$ and $B^0,B^1\in\Qmin^{n\times n}$ such that, if there exists a place $p$ with initial marking $\mu\in\{0,1\}$, upstream transition $t_j$ and downstream transition $t_i$, then $A^\mu_{ij} = \tau_p^-$ and $B^\mu_{ij} = \tau_p^+$, otherwise $A^\mu_{ij} = -\infty$ and $B^\mu_{ij}=+\infty$.
Let $x_i(k)\in\Q$ denote the time when transition $t_i\in\transitions$ fires for the $k$-th time, where $i\in\dint{1,n}$ and $k\in\N$.
Since the $(k+1)$-st firing of any transition $t_i$ cannot occur before the $k$-th one, it is natural to assume that $x_i$ is nondecreasing in $k$, i.e., $x_i(k+1) \geq x_i(k)$ for all $i\in\dint{1,n}$ and $k\in\N$.
%The dater function $x:\No\rightarrow \R^n$ represents the firing time of transitions; element $x_i(k)$ is the time of the $(k+1)$st firing of transition $t_i$.
%Since the $k$th firing of any transition $t_i\in\transitions$ cannot occur before the $(k+1)$st, it is natural to assume that $x_i(k-1) \leq x_i(k)$ for all $k\in\N$.
The dynamics of a P-TEG can be described by the following system of infinitely many linear inequalities in infinitely many variables $x_i(k)$: for all $i,j\in\dint{1,n}$, $\mu\in\{0,1\}$, $k\in\N$,
%The evolution of a P-TEG can now be described by the following set of inequalities (for a more detailed explanation, see~\cite{vspavcek2021analysis}): $\forall k\in\No$, $i\in\{1,\ldots,n\}$,
\begin{equation}\label{eq:dynamics_PTEGs}
	\left\{
	\begin{array}{rcl}
	A^\mu_{ij} + x_j(k) \leq & x_i(k+\mu) & \leq B^\mu_{ij} + x_j(k),\\
	x_i(k) \leq & x_i(k+1). & 
	\end{array}
	\right.
\end{equation}
The meaning of the inequalities in the first line of~\eqref{eq:dynamics_PTEGs} is that, in order to satisfy the constraints imposed by the time interval $[\tau_p^-,\tau_p^+]$ associated to place $p$ with $m(p) = \mu$ initial tokens, the downstream transition $t_i$ of $p$ needs to fire for the $(k+\mu)$-th time at least $A_{ij}^\mu = \tau_p^-$ time units and at most $B_{ij}^\mu = \tau_p^+$ time units after the $k$-th firing of the upstream transition $t_j$ of $p$.
The second line of~\eqref{eq:dynamics_PTEGs} simply imposes the nondecreasingness condition on $x_i$.
%For a more detailed explanation of the dynamical inequalities~\eqref{eq:dynamics_PTEGs}, the reader is referred to~\cite{vspavcek2021analysis}.
Note that matrices $A^0,A^1,B^0,B^1$ uniquely define a P-TEG with at most one initial token per place.
For this reason, they are called \emph{characteristic matrices} of the associated P-TEG.

\subsection{Consistency and $\N$-periodic graphs}

We can now give a more formal definition of consistency: a P-TEG is consistent if there exists an infinite trajectory $\{x_i(k)\in\Q\mid i\in\dint{1,n},\ k\in\N\}$ that satisfies~\eqref{eq:dynamics_PTEGs}.
Such a trajectory is then called consistent for the P-TEG, as it corresponds to an evolution of the marking in the P-TEG for which no token death occurs.
In this section, we consider the following decision problem.

\noindent
P-TEG CONSISTENCY\\
\textbf{Instance:} Matrices $A^0,A^1\in\Qmax^{n\times n}$, $B^0,B^1\in\Qmin^{n\times n}$.\\
\textbf{Question:} Is the P-TEG with characteristic matrices $A^0$, $A^1$, $B^0$, $B^1$ consistent?

To reduce the number of matrices involved, it is worth stating~\eqref{eq:dynamics_PTEGs} in the following, equivalent form
\begin{equation}\label{eq:simpler_dynamics}
	\begin{array}{l}
		\forall k\in\N,\\
		\forall i,j\in\dint{1,n},\\
	\end{array}
	\quad
	\left\{
	\begin{array}{rl}
		x_i(k) \geq & (M_{-1})_{ij} + x_j(k+1),\\
		x_i(k) \geq & (M_0)_{ij} + x_j(k),\\
		x_i(k+1) \geq & (M_{+1})_{ij} + x_j(k),
	\end{array}
	\right.
\end{equation}
where we used matrices $M_{-1},M_0,M_{+1}\in\Qmax^{n\times n}$ defined by $(M_0)_{ij} = \max(A^0_{ij},\,-B^0_{ji})$,
	$(M_{+1})_{ij} = A^1_{ij}$ if $i\neq j$, $(M_{+1})_{ij}=\max(0,\, A^1_{ii})$ if $i=j$,
%	\begin{dcases}
%		A^1_{ij} & \mbox{if } i\neq j,\\
%		\max(0,\, A^1_{ii}) & \mbox{if } i = j,
%	\end{dcases}
and $(M_{-1})_{ij} = -B^1_{ji}$.
The equivalence between~\eqref{eq:dynamics_PTEGs} and~\eqref{eq:simpler_dynamics} can be easily verified using the fact that, for any $x,y,a,b\in\Q$, the inequalities $x \geq a + y$ and $x\geq b + y$ hold if and only if $x \geq \max(a + y,b + y) = \max(a,b)+y$.
Even more compactly, we can write,\rerewrite{ $\forall k\in\N$, $i\in\dint{1,n}$,
\[
	\left\{
	\begin{array}{rl}
		x_i(k) \geq & \displaystyle\max_{j\in\dint{1,n}}\left\{(M_{-1})_{ij} + x_j(k+1)\right\},\\
		x_i(k) \geq & \displaystyle\max_{j\in\dint{1,n}}\left\{(M_0)_{ij} + x_j(k)\right\},\\
		x_i(k+1) \geq & \displaystyle\max_{j\in\dint{1,n}}\left\{(M_{+1})_{ij} + x_j(k)\right\},
	\end{array}
	\right.
\]
or,}{} using the max-plus algebra,
\begin{equation}\label{eq:simple_dynamics_PTEGs}
	\forall k\in\N,
	\quad
	\left\{
	\begin{array}{rl}
		x(k) \geq & M_{-1} \otimes x(k+1),\\
		x(k) \geq & M_0 \otimes x(k),\\
		x(k+1) \geq & M_{+1} \otimes x(k).
	\end{array}
	\right.
\end{equation}

Define $x = [x(1)^\top\ x(2)^\top \cdots]^\top\in\Q^\N$ and $M\in\Qmax^{\N\times \N}$ as in~\eqref{eq:block_tridiagonal}.
Then, \eqref{eq:simple_dynamics_PTEGs} can be rewritten as $x\geq M\otimes x$.
Therefore, the P-TEG corresponding to matrices $M_{-1},M_{0},M_{+1}$ is consistent if and only if $x\geq M\otimes x$ admits a rational solution $x\in\Q^\N$.
From Proposition~\ref{pr:infinite_matrix_inequalities}, the existence of a rational solution is equivalent to $\graph(M)$ not containing $\infty$-weight paths.
Recalling that $\graph(M)$ is isomorphic to an $\N$-periodic graph, Theorem~\ref{th:polynomial_complexity} implies that it is possible to solve the problem P-TEG CONSISTENCY in strongly polynomial time $\pazocal{O}(n^6)$.

%\begin{rem}
%Note that, if the initial number of tokens in each place of the P-TEG is encoded in binary, the complexity of the algorithm is pseudo-polynomial when applied to P-TEGs in which at least one place initially contains more than one token.
%This is due to the transformation recalled in Remark~\ref{re:transformation}, which, according to~\eqref{eq:transform_PTEG}, increases the number of transitions in a polynomial way with respect to the number of initial tokens in the Petri net.
%Considering a unary encoding of $m(p)$ for all $p\in\places$, the time complexity of the algorithm remains strongly polynomial.
%\myenddefinitionsymbol
%\end{rem}

\begin{exmp}
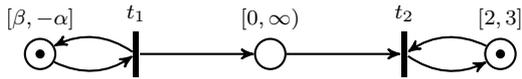
\begin{figure}
	\centering
	%\resizebox{1\linewidth}{!}{
	\begin{tikzpicture}[node distance=.5cm and 1.5cm,>=stealth',bend angle=30,thick]
\footnotesize
%\node[transitionV,label=below:{$t_1$}] (t1) {};
%\node[place,right=of t1,label=above:{$[0,\gamma_\wZ]$}] (p12) {};
%\node[transitionV,right=of p12,label=below:{$t_2$}] (t2) {};
%\node[place,tokens=1,above=of t1,label=above:{$[\alpha_\wZ,\alpha_\wZ]$}] (p11) {};
%\node[place,tokens=1,above=of t2,label=above:{$[\beta_\wZ,\beta_\wZ]$}] (p22) {};
%
%\draw (t1) edge[->] (p12);
%\draw (p12) edge[->] (t2);
%\draw (t1.90-15) edge[bend right,->] (p11);
%\draw (p11) edge[bend right,->] (t1.90+15);
%\draw (t2.90-15) edge[bend right,->] (p22);
%\draw (p22) edge[bend right,->] (t2.90+15);

\node[transitionV,label=above:{$t_1$}] (t1) {};
\node[place,right=of t1,label=above:{$[0,\infty)$}] (p12) {};
\node[transitionV,right=of p12,label=above:{$t_2$}] (t2) {};
\node[place,tokens=1,left= 1cm of t1,label=above:{$[\beta,-\alpha]$}] (p11) {};
\node[place,tokens=1,right= 1cm of t2,label=above:{$[2,3]$}] (p22) {};

\draw (t1) edge[->] (p12);
\draw (p12) edge[->] (t2);
\draw (t1) edge[bend right,->] (p11);
\draw (p11) edge[bend right,->] (t1);
\draw (t2) edge[bend right,->] (p22);
\draw (p22) edge[bend right,->] (t2);

\end{tikzpicture}
	%}
	\caption{Example of P-TEG.}
	\label{fi:P-TEG_example}
\end{figure}

Consider the P-TEG represented in Figure~\ref{fi:P-TEG_example}.
Its characteristic matrices are
	$A^0 = \begin{bsmallmatrix}
	-\infty & -\infty \\
	0 & -\infty
\end{bsmallmatrix}$,
	$A^1 = \begin{bsmallmatrix}
	\beta & -\infty \\
	-\infty & 2
\end{bsmallmatrix}$,
	$B^0 = \begin{bsmallmatrix}\infty & \infty\\\infty & \infty\end{bsmallmatrix}$,
	$B^1 = \begin{bsmallmatrix}
	-\alpha & \infty \\
	\infty & 3
\end{bsmallmatrix}$, 
where $\alpha\leq 0$, $\beta\geq 0$ are some parameters.
By applying the formulas for $M_{-1},M_0,M_{+1}$ from the previous subsection, we get 
	$M_{-1}=-B^{1\top}=\begin{bsmallmatrix}
	\alpha & -\infty\\-\infty & -3
	\end{bsmallmatrix}$,
	$M_{+1} = A^1\oplus E = \begin{bsmallmatrix}
	\beta & -\infty\\-\infty & 2
\end{bsmallmatrix}$, $M_0 = A^0\oplus (-B^{0\top})=
	\begin{bsmallmatrix}
		-\infty & -\infty \\ 0 & -\infty
	\end{bsmallmatrix}$.
Note that these matrices coincide with the ones from Example~\ref{ex:simple_graph}.
Therefore, from the discussion of Example~\ref{ex:periodic_graphs_max_plus}, we can conclude that the P-TEGs obtained by setting $\alpha=-1,\beta=2$ and $\alpha=-5,\beta=4$ are not consistent, and the one corresponding to values $\alpha=-1,\beta=1$ is consistent.
It can indeed be verified that the trajectory
	$x(1) = \begin{bsmallmatrix}
		0\\0
	\end{bsmallmatrix}$,
	$x(k+1) = \begin{bsmallmatrix}
		1\\2
	\end{bsmallmatrix} + x(k)$ $\forall k\in\N$
satisfies~\eqref{eq:simpler_dynamics}.
\myenddefinitionsymbol
\end{exmp}
\rewrite{\section{Conclusions}

We proposed an algorithm that verifies the consistency property in P-TEGs with $n$ transitions and at most one initial token per place in time $\pazocal{O}(n^6)$.
On a technical side, assuming binary encoding of the number of initial tokens per place $m(p)$, this shows that consistency can be checked in pseudo-polynomial time in a P-TEG with $m(p)>1$ for some place $p$, since the transformation outlined in Remark~\ref{re:transformation} adds transitions in polynomial number with respect to $m(p)$.
With unary encoding of $m(p)$, the time-complexity is instead polynomial.

Before addressing questions related to performance evaluation and control of P-TEGs, one last technical problem remains to be solved: the computation of the largest weight of all paths from a source to a target node in $\N$-periodic graphs.
We leave this problem for future work.}{}{}

\begin{ack}
The first author would like to thank Stephane Gaubert and Laurent Hardouin for the fruitful discussions.
\end{ack}

\bibliography{references}
\end{document}